\documentclass[]{jsedi_sub}
\title{Large-scale bioacoustic detection using semantic segmentation: a deep learning framework applied to fin whale calls in ocean-bottom seismometer recordings}

\author[1]{Jocelyn Japnanto
	\orcid{0009-0008-0912-4065}
	\thanks{Corresponding author: 
\href{mailto:zcapjja@ucl.ac.uk}{\texttt{jocelyn.japnanto.20@ucl.ac.uk}}}
}
\author[1,2]{Alex A. Saoulis
	\orcid{0009-0005-1486-8681}
}
\author[3,4]{Miriam Romagosa
	\orcid{0000-0003-2781-5528}
}
\author[3]{Rita Leitão
	\orcid{0009-0002-4979-1408}
}
\author[5]{Gabrielle Arrieta
	\orcid{0009-0003-7470-5345}
}
\author[3,6]{Mónica A. Silva
	 \orcid{0000-0002-2683-309X}
}
\author[7]{Matthew Graham
	\orcid{0000-0001-9104-7960}
}
\author[1]{Ana M. G. Ferreira
	\orcid{0000-0002-9492-6415}
}
\affil[1]{Department of Earth Sciences, University College London, London, United Kingdom}
\affil[2]{Department of Physics and Astronomy, University College London, London, United Kingdom}
\affil[3]{Instituto de Investigação em Ciências do Mar (OKEANOS), Universidade dos Açores, Horta, Portugal}
\affil[4]{Marine Geosciences Department, Institute of Marine Sciences, CSIC, 08003 Barcelona, Spain}
\affil[5]{Centre for Research into Ecological \& Environmental Modelling, University of St Andrews, St Andrews, United Kingdom}
\affil[6]{Instituto do Mar (IMAR), Horta, Portugal}
\affil[7]{Advanced Research Computing Centre, University College London, London, United Kingdom}

\usepackage{multirow}
\usepackage{tabularx}

\newcolumntype{C}{>{\centering\arraybackslash}X} 
\usepackage{makecell}

\usepackage{array}
\usepackage{colortbl}

\usepackage[percent]{overpic} 
\usepackage{subcaption}

\begin{document}

\publicationonly{
\dois{}
\handedname{Efirstname Elastname}
\receiveddate{September 23, 2024}
\accepteddate{January 6, 2025}
\publisheddate{January 11, 2025}
\theyear{2025}
\thevolume{2}
\thepaper{1}  
}

\addsummaries{
  \begin{summary}{Abstract}
  Ocean-bottom seismometers (OBS), originally deployed for geophysical research, continuously record low-frequency sound for months to years across broad areas of ocean, offering a largely untapped resource for passive acoustic monitoring (PAM) of baleen whales. Realising this potential requires automated detection methods that operate reliably across the varied recording conditions found in large sensor networks. We present a deep learning semantic segmentation framework that detects the 20-Hz notes of fin whales (\textit{Balaenoptera physalus}) in OBS spectrograms, assigning each pixel a probability of belonging to a call and converting the resulting probability maps into time-frequency bounding boxes describing individual detections. We trained the model on hydrophone recordings from one OBS deployment in the Azores-Madeira-Canaries region and applied it without retraining to vertical-component seismometer recordings from a second, geographically distinct deployment, showing that a single trained model generalises across sensor types and recording environments. Applied to 378,912~hours of recordings from 46 OBS sites, the detector identified 6.3~million calls, forming the largest fin whale call catalogue assembled to date, with high precision ($\sim97$\%) across both deployments. The resulting catalogue resolves call timing and spectral structure accurately enough to support ecological analyses, revealing coherent seasonal shifts in three persistent inter-note interval (INI) groups across the singing season and basin-scale patterns in calling activity. By transforming existing geophysical infrastructure into a scalable sensing network, our approach substantially expands the spatial and temporal reach of PAM without new hardware investment, offering a transferable framework for tracking other low-frequency vocalising species and informing conservation planning, marine spatial management, and abundance estimation efforts across large scales.
  \end{summary}
  }

\begin{summary}{Key words}
    \textit{Bioacoustics, deep learning, seismometer, passive acoustic monitoring}
    \end{summary}

\section{Introduction}
Baleen whales are among the most wide-ranging mammals on Earth, undertaking seasonal movements that challenge conventional monitoring approaches. Because their distribution, population dynamics, and behavioural patterns operate across vast spatial and temporal extents, effective conservation management depends on monitoring tools that can match these ecological scales. Visual surveys provide valuable data but are constrained by weather, daylight, and logistical limitations. Tagging and genetic sampling yield detailed individual-level information but are difficult to sustain across the broad spatial extents and continuous time periods needed to track population-level patterns \parencite{thomas2012passive}. Conventional monitoring methods thus lack both the spatial coverage and the temporal continuity that population-level monitoring requires.

PAM offers a powerful complement to these methods. Many baleen whales produce stereotyped, low-frequency vocalisations that propagate over long distances, enabling detection across vast areas and during periods when visual surveys are not feasible \parencite{mcdonald1999passive}. Continuous recordings can provide uninterrupted coverage over months to years, but the resulting data volumes make manual analysis impractical at scale \parencite{van2009management, stowell2022computational, kershenbaum2025automatic}, making automated detection methods essential for realising the full potential of long-duration acoustic datasets.

Fin whales (\textit{Balaenoptera physalus}) produce stereotyped, high-amplitude pulses around 20~Hz in repetitive sequences forming songs that last minutes to hours and are widely thought to serve a reproductive or social function \parencite{watkins198720, croll2002only}. The timing between successive calls, quantified as the INI, describes song structure and varies across ocean basins. INIs have been shown to change both abruptly and gradually over time, with studies in the North Atlantic and the Mediterranean documenting long-term shifts in dominant INI values \parencite{romagosa2024fin, guazzo2024decade, best2022temporal}. Intra-seasonal INI variation has been reported in the North Pacific  \parencite{oleson2014synchronous} and the Northwest Atlantic \parencite{morano2012seasonal}, suggesting that song timing is a dynamic rather than fixed property of fin whale populations. Seasonal variation in calling activity is also well-established, with elevated call rates typically observed from late autumn through early spring in the North Atlantic \parencite{nieukirk2004low, romagosa2020baleen}.

OBSs are designed for geophysical research but also record low-frequency acoustic signals that couple into the seafloor, routinely capturing fin whale calls \parencite[e.g.,][]{mcdonald1995blue, wilcock2012tracking, pereira2021source}. OBS deployments record continuously for months to over a year and can span large spatial areas, providing the kind of broad-scale, long-duration coverage that dedicated hydrophone arrays rarely achieve \parencite[e.g.,][]{Barruol2013, Ferreira2026}. While OBS data have been used for ecological studies including seasonal presence monitoring, whale tracking, density estimation, and assessment of anthropogenic noise impacts \parencite[e.g.,][]{hilmo2025applying, edwards2026seismic}, most prior work has focused on only a few instruments or short time windows. The full spatial and temporal extent of large network deployments remains largely unexploited for ecological purposes.

Machine learning has become a popular approach for analysing large acoustic datasets \parencite[e.g.,][]{stowell2022computational, kershenbaum2025automatic, allen2021convolutional, bergler2019orca}. Template-based detectors can perform well under stable conditions but require manual tuning and degrade when signal characteristics or noise conditions vary across sensors or deployments \parencite{mellinger1997methods}. Convolutional neural networks applied to spectrograms learn relevant features directly from labelled data, making them more robust to varying noise conditions and signal characteristics than template-based methods \parencite{kershenbaum2025automatic}. This robustness is particularly valuable for large-scale deployments, where recording conditions can vary substantially across instruments and sites, as has been demonstrated for cetacean call detection across diverse long-term datasets \parencite{allen2021convolutional}. Most existing systems produce event-level outputs indicating vocalisation presence within a fixed time window \parencite{stowell2022computational}, which limits the information available about call structure. We address this by formulating detection as a semantic segmentation problem in the time-frequency domain, in which the model assigns a probability to every pixel in a spectrogram, yielding spectrally and temporally resolved detections that allow direct estimation of call timing and frequency bounds. This approach builds on the segmentation framework of \textcite{saoulis2026semantic}, which demonstrated strong performance for detecting blue whale calls in OBS spectrograms, and extends it to fin whale 20-Hz notes across a large deployment.

We develop and evaluate a supervised segmentation-based detector for 20-Hz notes in OBS spectrograms, trained on hydrophone recordings from one site in the central Azores, and applied without retraining to a large regional seismometer network spanning the Azores-Madeira-Canaries region. We use this detector to produce the largest continuous catalogue of fin whale calls to date across the eastern North Atlantic, demonstrating the feasibility of automated PAM at basin scale using geophysical instruments. We then characterise seasonal patterns in call occurrence and INI variation across the array, illustrating the ecological information that can be extracted from OBS datasets through automated deep learning methods.

\section{Materials and Methods}
\subsection{Data collection}

\begin{figure}[h!]
  \includegraphics[width=\textwidth]{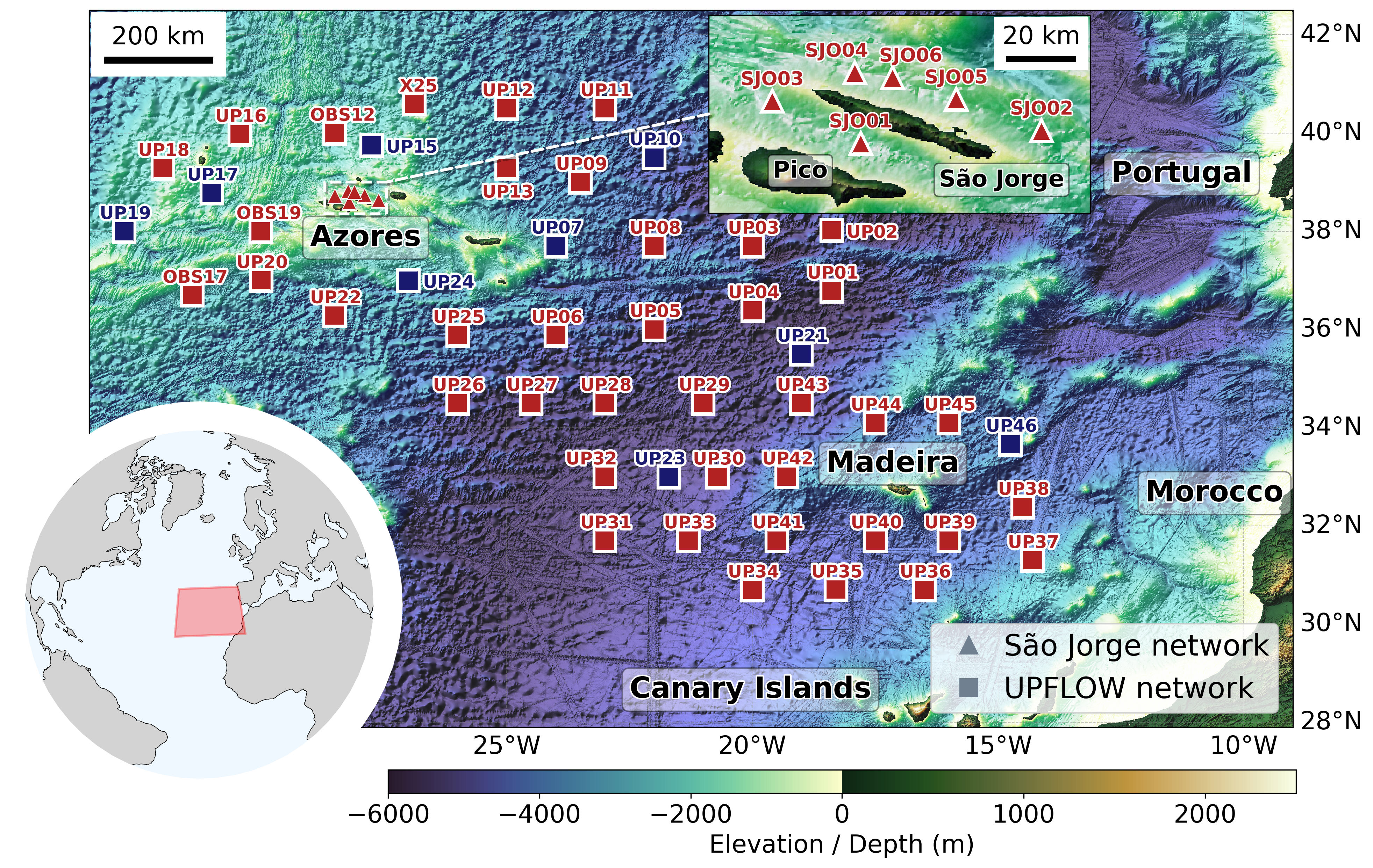}
  \caption{Map of the São Jorge (triangles) and UPFLOW (squares) OBS arrays located across the Azores-Madeira-Canaries region. Each symbol marks an individual OBS station, with those coloured red being used in this study. Hydrophone-component data from São Jorge (August 2022 to February 2023), and vertical-component data from UPFLOW (June 2021 and August 2022), were used to detect fin whale calls.}
  \label{fig:UPFLOW_SJ_Map}
\end{figure}

We use high-quality OBS data from two deployments in the North Atlantic Ocean: the São Jorge network \parencite{Ferreira2022} and the UPFLOW \parencite[\textbf{UP}ward mantle \textbf{FLOW} from novel seismic observations;][]{Ferreira2024, Tsekhmistrenko2026}
network (Fig. \ref{fig:UPFLOW_SJ_Map}). Both networks were designed for geophysical research, but their broadband seismometers and hydrophones also record low-frequency acoustic energy propagating through the water column and seafloor, making them useful for other fields such as marine bioacoustic and oceanographic studies \parencite[e.g.,][]{saoulis2026semantic}.

The São Jorge deployment features six short-period OBSs \parencite{minshull2005multi} deployed around São Jorge Island in the central Azores, operating from August 2022 to February 2023. The UPFLOW deployment formed a regional-scale network of 50 OBSs (49 recovered) across the Azores-Madeira-Canaries region across an area of $\sim 1,000 \times 2,000 \text{km}^{2}$, deployed between June 2021 and August 2022. 

We used data recorded by the hydrophone component for São Jorge and the vertical-component seismometer for UPFLOW, as these produced the most consistently well-defined 20-Hz notes within each deployment. We excluded nine UPFLOW stations due to insufficient sampling rates, hardware malfunction, and/or recording failure. The final dataset comprised 378,912 hours of recordings from 46 stations. Waveform data are deposited in the Earthscope and GFZ EIDA data centres under network codes 4U and 8J, respectively. Instrument specifications and recording characteristics for all stations are provided in Table~\ref{SI: station table}. 

\subsection{Data processing}

\begin{figure}[h!]
    \begin{flushright}
      \includegraphics[width=15cm,
      trim={0 0 0 1.4cm}, clip
      ]
      {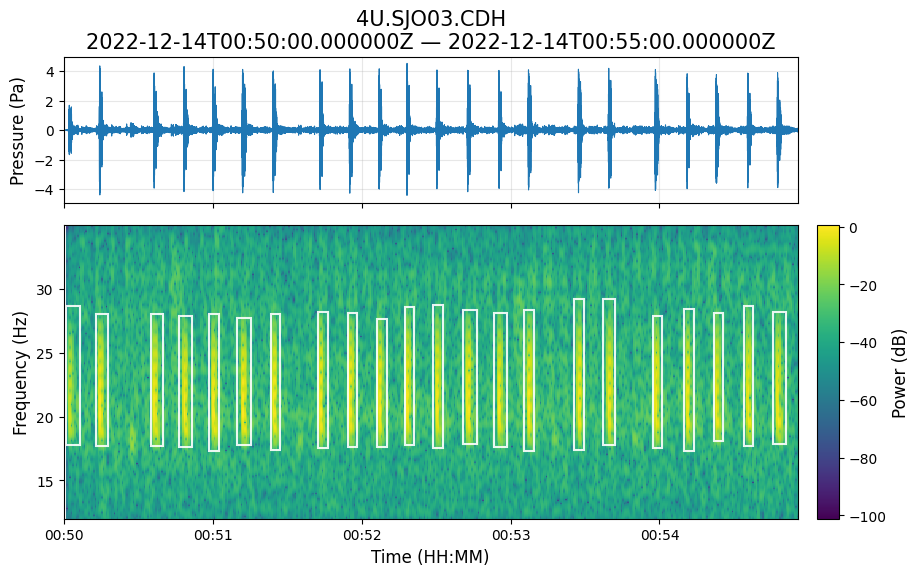}
      \put(-450,220){\large a)}   
      \put(-450,145){\large b)}   
    
      \caption{A 5-minute recording showing part of a fin whale song sequence at station SJO03, located west of the São Jorge Island. \textbf{a)} Waveform time series and \textbf{b)} spectrogram of the hydrophone component.  Instrument response was removed and the trace was band-pass filtered between 12 to 35~Hz with a Butterworth filter. The spectrogram was calculated using a short-time Fourier transform with a window length of 2.0~s and 75\% overlap. Manual annotations of 20-Hz notes are shown as white boxes. }
      \label{fig:finwhale_call_wavefrom_spectrogram_SJO03_CHZ}
    \end{flushright}
\end{figure}

We segmented each recording into non-overlapping 5-minute windows, a duration that captures up to $\sim$30 consecutive calls from a single individual (Fig.~\ref{fig:finwhale_call_wavefrom_spectrogram_SJO03_CHZ}). For each window we computed a spectrogram using a short-time Fourier transform with a 2.0~s Hanning window and 75\% overlap, chosen to balance the temporal resolution needed to separate successive calls against the frequency resolution needed to characterise their spectral content. We restricted each spectrogram to the 12-35~Hz band, represented it as a single-channel greyscale image, and rescaled it to $480\times640$~px from its original $369\times496$~px to meet the input requirements of the segmentation architecture.


\begin{figure}[h!]
      \includegraphics[width=15cm,
      ]
      {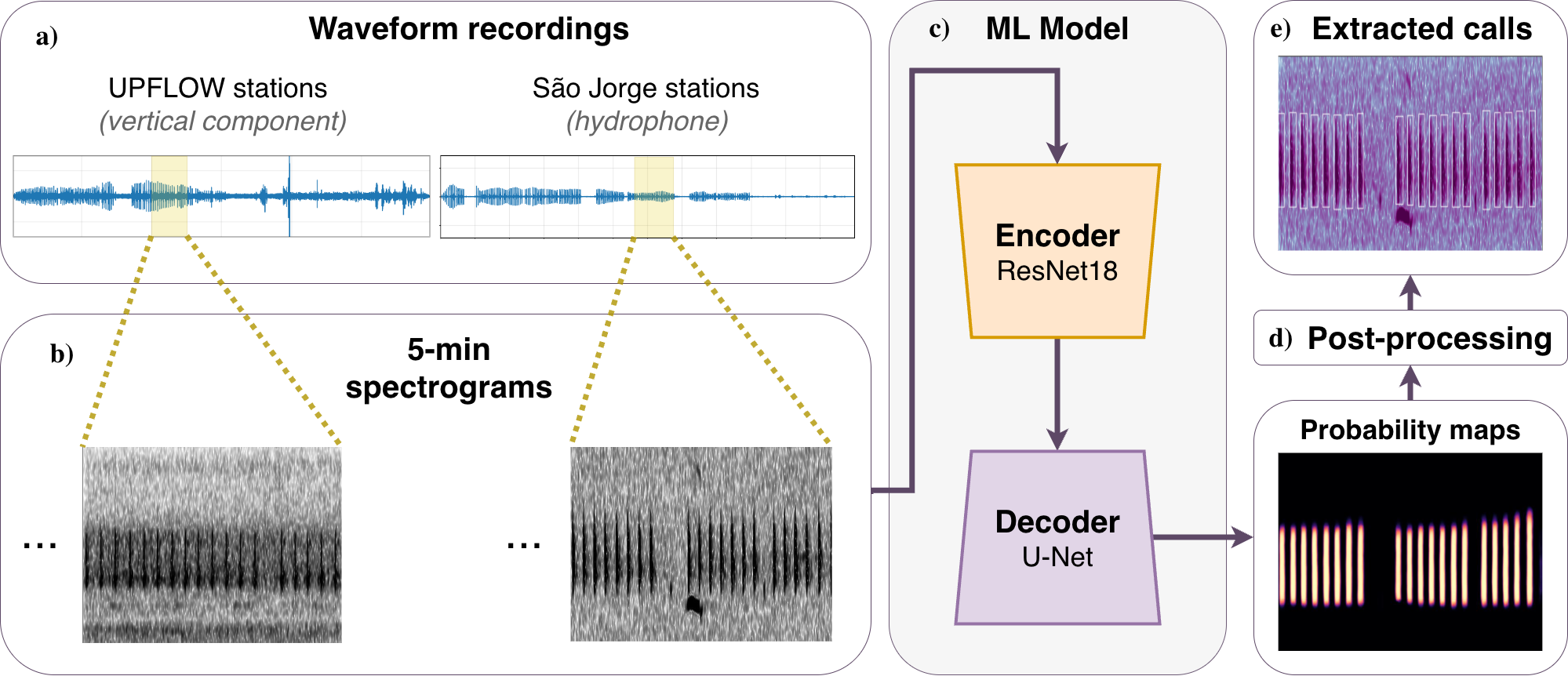}
    
      \caption{Overview of the data processing and detection pipeline. \textbf{a)} Raw waveforms are segmented into five-minute windows. \textbf{b)} Greyscale spectrograms, computed via short-time Fourier transforms, serve as input to the model.  \textbf{c)} The encoder-decoder machine learning (ML) model. \textbf{d)} Post-processing extracts individual calls from the model's predictions.  \textbf{e)} Resulting detections (white boxes) form the call catalogue.}
      \label{fig:Detection_Pipeline}
\end{figure}

For training spectrograms, we generated binary label masks from the annotation catalogue, marking annotated calls as foreground pixels. Fig.~\ref{fig:Detection_Pipeline} gives an overview of the pre-processing and detection pipeline.

\subsection{Training dataset}

Expert analysts manually annotated 20-Hz notes in hydrophone recordings from station SJO03 using Raven Pro v. 1.6.5 \parencite{ravenpro165}, focusing on periods of high call density during October and December 2022. This produced an annotated catalogue of nearly 40,000 calls.

To promote robustness against false detections across diverse recording environments, background samples containing no fin whale calls were also included, drawn from stations SJO03, UP36, and UP41 across a total of nine days of manually verified recordings. The final dataset comprised approximately 4,200 five-minute spectrograms ($\sim$350 hours), of which roughly 40\% contained calls. Data were partitioned into training (70\%), validation (15\%), and testing (15\%) subsets using stratified sampling to preserve the proportion of call-positive and call-negative samples across subsets. Representative annotated and background samples are shown in Fig.~\ref{fig:training_samples}.

\subsection{Model architecture and training}

We framed detection as a semantic segmentation problem, in which the model predicts a class label for each pixel in the input image, distinguishing calls from background. This pixel-level inference yields fine-grained detections rather than the window-level presence or absence reported by classification-based detectors.

Building on previous work from \textcite{saoulis2026semantic}, our model follows a encoder-decoder architecture. The encoder extracts features from the input spectrogram while progressively reducing their spatial dimensions. The decoder then progressively upsamples these low-resolution feature maps back to the original input resolution to produce a dense, pixel-wise prediction. Feature extraction is performed using a ResNet-18 backbone \parencite{he2016deep} pretrained on the ImageNet dataset \parencite{deng2009imagenet}. Although ImageNet contains natural photographs rather than spectrograms, the early convolutional layers of deep networks trained on large image datasets learn generalisable low-level features including edges, contours, and textures. These features have been shown to provide a useful initialisation for similar detection tasks \parencite[e.g.,][]{fonseca2021analysis,mao2022automated}. The decoder follows a U-Net style architecture \parencite{ronneberger2015u}, in which skip connections pass high-resolution spatial feature maps from each encoder stage directly to the corresponding decoder stage. This allows the network to recover fine-scale temporal and spectral details during upsampling that would otherwise be lost through downsampling in the encoder. The final output is a dense probability map of the same spatial dimensions as the input spectrogram, with each pixel assigned a value between 0 and 1 representing the probability that it belongs to a 20-Hz note. Full details of the model architecture are provided in \textcite{saoulis2026semantic}.

We trained the model in mini-batches of 12, minimising the binary cross-entropy loss between predicted probability maps and ground-truth masks using the \textit{Adam} optimiser \parencite{kingma2014adam} with a learning rate of 0.0005 and weight decay factor of 0.01. Adam is a gradient-based optimisation algorithm that adapts the learning rate for each parameter from running estimates of the gradient's first and second moments, which speeds and stabilises convergence. Weight decay is a regularisation term that penalises large weights during training to discourage overfitting \parencite{krogh1991simple}. Training proceeded for a maximum of 50 epochs, with early stopping applied when validation loss failed to improve for five consecutive epochs. The best-performing model checkpoint, selected on the basis of lowest validation loss, was retained for inference. All experiments were conducted with a fixed random seed to ensure reproducibility of dataset splits and parameter initialisation. We then applied the trained model to the remaining spectrograms and post-processed the resulting probability maps into discrete detections.

Inference was performed on a workstation with an NVIDIA RTX A6000 GPU (48 GB VRAM) and a dual Intel Xeon Gold 6230 CPU (40 cores/80 threads, 187 GB RAM), using PyTorch 2.0.1 with CUDA 11.7. Once spectrograms had been computed, processing one full day of recordings from a single station, including model inference and post-processing, took approximately 15 to 20~s.

\subsection{Post-segmentation processing}\label{subsection: Post-processing}

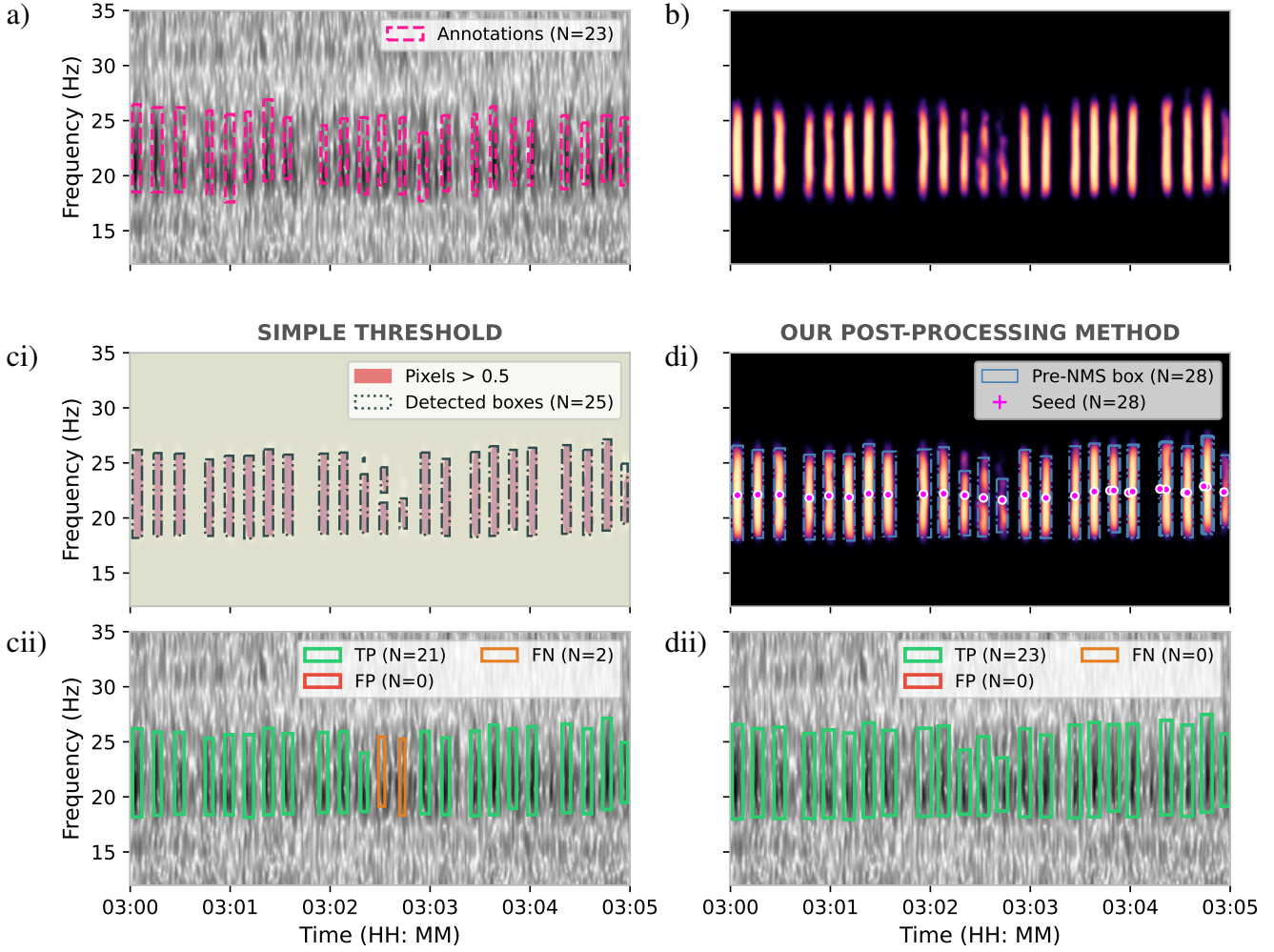
\begin{figure}[h!]
    \begin{flushright}
      \includegraphics[width=\textwidth,
      trim={0 0 0 0}, clip
      ]
      {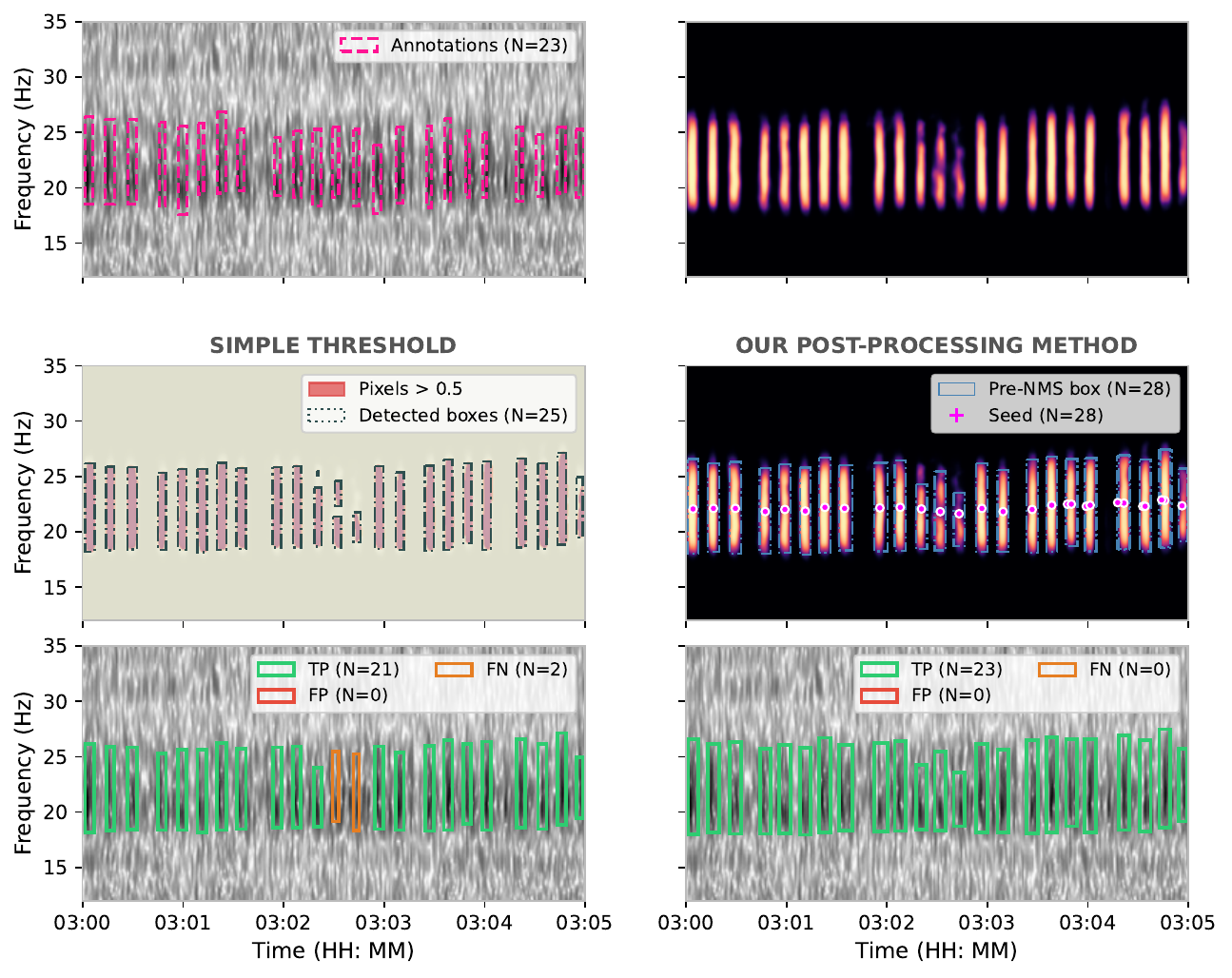}
      \put(-500,370){\large a)}   
      \put(-240,370){\large b)}     
      \put(-500,235){\large ci)}
      \put(-240,235){\large di)}   
      \put(-500,122){\large cii)}
      \put(-240,122){\large dii)}     
      \caption{Comparison of post-processing approaches to extract calls, applied to a 5-minute sample from the test dataset from station SJO03 (05-Dec-2022). \textbf{a)} Manual annotations of 20-Hz notes (pink). \textbf{b)} Corresponding 20-Hz note probability map, inferred by the detector. \textbf{ci)} A simple thresholding approach, where pixels exceeding a fixed probability threshold are grouped into connected components to form detections, \textbf{cii)} and the resulting calls after filtering, evaluated against annotations, showing true positives (TP; green), false positives (FP; red), and false negatives (FN; orange). \textbf{di)} Our method, where local maxima (“seeds”) in a smoothed response map initialise the formation of our candidate bounding boxes prior to non-maximum suppression (NMS), \textbf{dii)} and the resulting detections, processed and evaluated similarly. Around 03:02:30, simple thresholding fragments and discards noisy calls that our method recovers.
      }

      \label{fig:Post_processing}
    \end{flushright}
\end{figure}

The post-segmentation pipeline converts the dense probability map into discrete bounding boxes describing individual call detections (Fig. \ref{fig:Post_processing}). Directly thresholding the probability map, which assigns each pixel to either the call or background class according to a fixed probability cut-off, was found to produce fragmented contours at times, particularly for low-amplitude calls, making reliable extraction of individual calls difficult (Fig.~\ref{fig:Post_processing}ci). We therefore adopt a region-growing approach loosely inspired by the seeded region growing algorithm of \textcite{adams1994seeded}.

A $20 \times 200$~px template is convolved with the probability map, followed by max-pooling with a kernel of the same dimensions and upsampling to the original resolution with nearest-neighbour interpolation. Candidate seeds are identified as local maxima exceeding the probability threshold $T_{prob}$. From each seed, the detection region expands iteratively in four directions. Expansion in the vertical and horizontal directions continues while the mean probability of adjacent pixels exceeds the thresholds $T_{t,b}$ and $T_{t,r}$, respectively, yielding a rectangular bounding box that approximates the full temporal and spectral extent of the call. We recorded the mean foreground probability within each box as a confidence score, then applied non-maximum suppression \parencite{girshick2014rich}, retaining the highest-confidence box and suppressing any box whose intersection-over-union (IoU) with it exceeded 0.15 (Fig.~\ref{fig:Post_processing}dii). We selected all threshold parameters by grid search on a held-out dataset, quantifying performance using F1-score and mean IoU (Section~\ref{subsubsection: Matching procedure}). Finally, we applied liberal filtering criteria (Eq.~\ref{eq: filtering criteria}) to remove detections inconsistent with the typical spectral and temporal properties of 20-Hz notes. A full description of the bounding-box notation and matching procedure is given in Section~\ref{subsubsection: Matching procedure}.


\subsection{Evaluation metrics}

We assessed detector performance against two independent sets of manual annotations. The first was a held-out test subset from station SJO03, annotated by the same analysts who produced the training labels. The second comprised independent annotations from 16 spatially and temporally balanced UPFLOW stations, produced by a different annotator to evaluate cross-sensor generalisation.

We summarised detection performance using precision, recall, F1-score, and mean IoU (mIoU), and characterised localisation accuracy using mean absolute error in call onset time ($\mathrm{MAE}_{\mathrm{onset}}$), duration ($\mathrm{MAE}_{\mathrm{D}}$), and lower and upper frequency bounds ($\mathrm{MAE}_{\mathrm{f_{low}}}$, $\mathrm{MAE}_{\mathrm{f_{high}}}$), with systematic bias also quantified for onset time ($\mathrm{Bias}_{\mathrm{onset}}$). Full metric definitions are provided in Section~\ref{subsubsection: Detection metrics}.

\subsection{INI analysis}

To characterise temporal calling behaviour, INIs were computed as the difference between the start times of successive calls. Intervals exceeding 25~s were excluded to remove extended silent periods between bouts, or potential missed calls. Weekly summaries of INI statistics were then used to explore temporal variability in calling behaviour across stations and seasons. We also identified recurrent interval groups from the resulting INI distributions and tracked their relative composition over the singing season.

\newpage
\section{Results} \label{Results}
Our detector produced a combined catalogue of 6,291,626 calls across both deployments (Table~\ref{table: station detection summary}), comprising of 1,153,823 calls at the São Jorge array and 5,137,803 at the UPFLOW array.

\begin{table}[p]
\begin{tabularx}{\textwidth}{|C|C|X|X|}
\hline

\textbf{Network} &
\textbf{Station} &
\textbf{Total recording hours} &
\textbf{Number of detections} \\ \hline

\multirow{7}{*}{São Jorge} & SJO01                              & 3,792                 & 58,478               \\ \cline{2-4}
                           & SJO02                              & 3,792                 & 248,155              \\ \cline{2-4}
                           & SJO03                              & 3,768                 & 197,509              \\ \cline{2-4}
                           & SJO04                              & 3,768                 & 217,934              \\ \cline{2-4}
                           & SJO05                              & 3,768                 & 215,590              \\ \cline{2-4}
                           & SJO06                              & 3,768                 & 216,157              \\ \cline{2-4}
                           & \textbf{São Jorge Total}           & \textbf{22,656}       & \textbf{1,153,823}   \\
                           \hline
\multirow{41}{*}{UPFLOW}   & OBS12                              & 8,376                 & 151,637              \\ \cline{2-4}
                           & OBS17                              & 4,680                 & 131,885              \\ \cline{2-4}
                           & OBS19                              & 5,760                 & 249,914              \\ \cline{2-4}
                           & UP01                               & 9,216                 & 49,764               \\ \cline{2-4}
                           & UP02                               & 9,216                 & 59,600               \\ \cline{2-4}
                           & UP03                               & 9,216                 & 67,766               \\ \cline{2-4}
                           & UP04                               & 9,240                 & 187,402              \\ \cline{2-4}
                           & UP05                               & 9,216                 & 101,559              \\ \cline{2-4}
                           & UP06                               & 9,216                 & 126,426              \\ \cline{2-4}
                           & UP08                               & 9,240                 & 126,605              \\ \cline{2-4}
                           & UP09                               & 9,216                 & 35,390               \\ \cline{2-4}
                           & UP11                               & 9,240                 & 45,175               \\ \cline{2-4}
                           & UP12                               & 9,240                 & 41,514               \\ \cline{2-4}
                           & UP13                               & 9,288                 & 89,496               \\ \cline{2-4}
                           & UP16                               & 9,264                 & 257,244              \\ \cline{2-4}
                           & UP18                               & 9,288                 & 56,667               \\ \cline{2-4}
                           & UP20                               & 9,336                 & 229,811              \\ \cline{2-4}
                           & UP22                               & 9,336                 & 115,019              \\ \cline{2-4} 
                           & UP25                               & 9,432                 & 152,890              \\ \cline{2-4}
                           & UP26                               & 5,448                 & 190,022              \\ \cline{2-4}
                           & UP27                               & 9,432                 & 159,562              \\ \cline{2-4}
                           & UP28                               & 8,088                 & 1,868                \\ \cline{2-4}
                           & UP29                               & 9,432                 & 191,634              \\ \cline{2-4}
                           & UP30                               & 9,312                 & 236,565              \\ \cline{2-4}
                           & UP31                               & 9,312                 & 191,324              \\ \cline{2-4}
                           & UP32                               & 9,336                 & 167,092              \\ \cline{2-4}
                           & UP33                               & 9,312                 & 291,416              \\ \cline{2-4}
                           & UP34                               & 9,312                 & 162,631              \\ \cline{2-4}
                           & UP35                               & 9,312                 & 235,380              \\ \cline{2-4}
                           & UP36                               & 9,336                 & 117,821              \\ \cline{2-4}
                           & UP37                               & 9,312                 & 49,120               \\ \cline{2-4}
                           & UP38                               & 9,336                 & 35,244               \\ \cline{2-4}
                           & UP39                               & 9,336                 & 80,203               \\ \cline{2-4}
                           & UP40                               & 9,312                 & 87,329               \\ \cline{2-4}
                           & UP41                               & 9,336                 & 164,384              \\ \cline{2-4}
                           & UP42                               & 9,336                 & 124,060              \\ \cline{2-4}
                           & UP43                               & 9,336                 & 48,187               \\ \cline{2-4}
                           & UP44                               & 9,336                 & 49,690               \\ \cline{2-4}
                           & UP45                               & 9,336                 & 188,437              \\ \cline{2-4}
                           & X25                                & 7,632                 & 90,070               \\ \cline{2-4}
                           & \textbf{UPFLOW Total}       & \textbf{356,256}      & \textbf{5,137,803}   \\ \hline
                           & \textbf{Total across all stations} & \textbf{378,912}      & \textbf{6,291,626}   \\ \hline
\end{tabularx} 

\caption{Number of 20-Hz notes detected at each station in the São Jorge and UPFLOW arrays.}
\label{table: station detection summary}

\end{table}

\subsection{Detector performance}

Detector performance is summarised in Table \ref{table:evaluation_results}. On the held-out São Jorge test set, the detector achieved precision of 94\%, recall of 80\%, and F1 of 87\% (mIoU = 61\%), with no false positives on background-only samples. Localisation accuracy was strong, with $\mathrm{MAE}_{\mathrm{onset}} = $ 0.98~s, $\mathrm{MAE}_{\mathrm{D}} = $ 2.02~s, and $\mathrm{MAE}_{\mathrm{f_{low}}}$ and $\mathrm{MAE}_{\mathrm{f_{high}}}$ both below 1~Hz.

Across the UPFLOW stations, precision remained consistently high (97\%, range 93-100\%), while recall was substantially lower (33\%, range 9-53\%), yielding an overall F1 of 49\%. Localisation errors were larger than at São Jorge ($\mathrm{MAE}_{\mathrm{onset}} = $ 2.86~s, $\mathrm{MAE}_{\mathrm{D}} = $ 7.53~s), though frequency estimates remained relatively accurate ($\mathrm{MAE}_{\mathrm{f_{low}}} = $ 0.70~Hz, $\mathrm{MAE}_{\mathrm{f_{high}}} =$ 1.42~Hz). 

\begin{figure}[h!]
    \begin{flushright}
      \includegraphics[width=0.98\textwidth, trim={0 31cm 0 1cm}, clip]{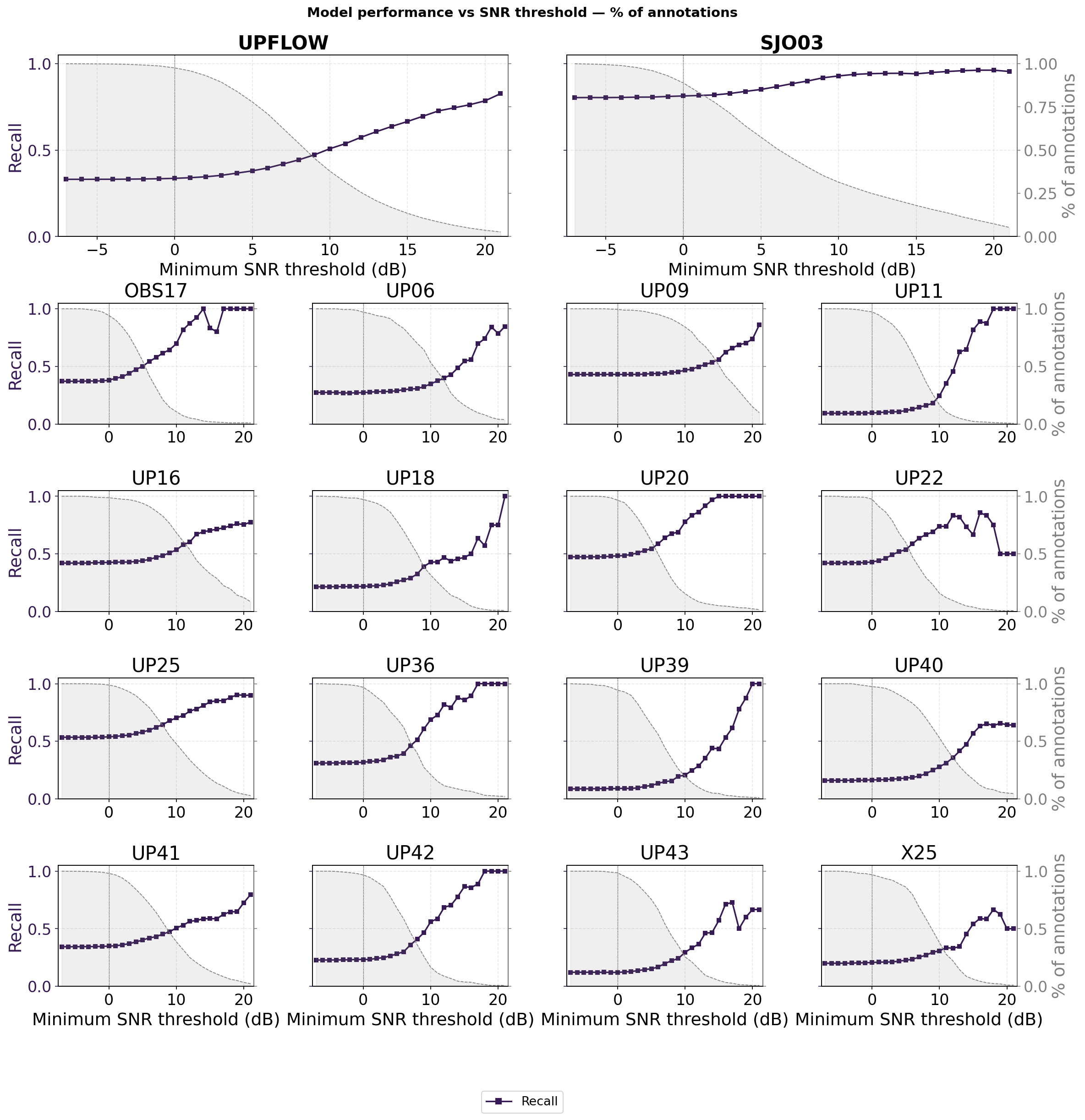}
      \put(-485,102){\large a)} 
      \put(-240,102){\large b)} 
      \caption{Recall as a function of minimum SNR threshold applied to \textbf{a)} an aggregation of all UPFLOW annotations, and \textbf{b)} the held-out test set from station SJO03. The threshold sets the minimum SNR an annotated call must exceed to enter the evaluation set, and the grey shading shows the number of annotations retained at each threshold.}
      \label{fig:SNR_recall}
    \end{flushright}
\end{figure}

Recall increased substantially with increasing SNR threshold applied to annotations (Fig. \ref{fig:SNR_recall}), confirming that missed detections are disproportionately concentrated among low-amplitude calls in the UPFLOW catalogue. Details on SNR definition and calculation can be found in Section~\ref{SI: SNR}. Station-level SNR-recall breakdowns are provided in Fig.~\ref{fig:SNR_recall_by_station}.

\clearpage
\thispagestyle{empty}

\begin{table}[p]
\centering

\begin{tabularx}{\textwidth}{|c|c|c|c|c|X|X|X|X|}
\hline
\textbf{Network}         & \textbf{Station} & \textbf{TP} & \textbf{FP} & \textbf{FN} & \textbf{Precision} & \textbf{Recall} & \textbf{F1 Score} & \textbf{mIoU} \\ \hline
São Jorge                & \textbf{SJO03}            & \textbf{4583}        & \textbf{267}         & \textbf{1137}        & \textbf{0.94}               & \textbf{0.80}            & \textbf{0.87}              & \textbf{0.61}              \\ \hline
\multirow{16}{*}{UPFLOW} & OBS17            & 114          & 2          & 196         & 0.98               & 0.37            & 0.53              & 0.13              \\ \cline{2-9}
                         & UP06             & 93          & 2           & 247         & 0.98               & 0.27            & 0.43              & 0.13              \\ \cline{2-9} 
                         & UP09             & 187         & 1           & 250         & 0.99               & 0.43            & 0.60              & 0.12              \\ \cline{2-9} 
                         & UP11             & 46          & 1           & 447         & 0.98               & 0.09            & 0.17              & 0.13              \\ \cline{2-9} 
                         & UP16             & 199         & 2           & 275         & 0.99               & 0.42            & 0.59              & 0.13              \\ \cline{2-9} 
                         & UP18             & 84          & 0           & 309         & 1.00               & 0.21            & 0.35              & 0.14              \\ \cline{2-9} 
                         & UP20             & 247         & 9           & 277         & 0.96               & 0.47            & 0.63              & 0.15              \\ \cline{2-9} 
                         & UP22             & 133         & 7           & 184         & 0.95               & 0.42            & 0.58              & 0.14              \\ \cline{2-9} 
                         & UP25             & 926         & 32          & 819         & 0.97               & 0.53            & 0.69              & 0.15              \\ \cline{2-9} 
                         & UP36             & 90          & 5           & 200         & 0.95               & 0.31            & 0.47              & 0.14              \\ \cline{2-9} 
                         & UP39             & 43          & 3           & 451         & 0.93               & 0.09            & 0.16              & 0.14              \\ \cline{2-9} 
                         & UP40             & 88          & 0           & 461         & 1.00               & 0.16            & 0.28              & 0.16              \\ \cline{2-9} 
                         & UP41             & 739         & 22          & 1414        & 0.97               & 0.34            & 0.51              & 0.13              \\ \cline{2-9} 
                         & UP42             & 91          & 4           & 308         & 0.96               & 0.23            & 0.37              & 0.15              \\ \cline{2-9} 
                         & UP43             & 50          & 4           & 361         & 0.93               & 0.12            & 0.22              & 0.13              \\ \cline{2-9} 
                         & X25              & 77          & 1           & 308         & 0.99               & 0.20            & 0.33              & 0.13              \\ \cline{2-9} 
                         & \textbf{Total}            & \textbf{3207}        & \textbf{95}          & \textbf{6507}        & \textbf{0.97}               & \textbf{0.33}            & \textbf{0.49}              & \textbf{0.14}              \\ \hline
\end{tabularx}

\vspace{0.6em}

\begin{tabularx}{\textwidth}{|c|c|X|X|X|X|X|}
\hline
\textbf{Network}         & \textbf{Station}  & \textbf{$\mathrm{MAE}_{\mathrm{onset}}$ (s)} & \textbf{$\mathrm{Bias}_{\mathrm{onset}}$ (s)} & \textbf{$\mathrm{MAE}_{\mathrm{D}}$ (s)} & \textbf{$\mathrm{MAE}_{\mathrm{f_{low}}}$ (Hz)} & \textbf{$\mathrm{MAE}_{\mathrm{f_{high}}}$ (Hz)} \\ \hline
São Jorge                & \textbf{SJO03}              & \textbf{0.98}                    & \textbf{-0.80}                   & \textbf{2.02}                      & \textbf{0.32}                        & \textbf{0.74}                         \\ \hline
\multirow{16}{*}{UPFLOW} & OBS17                       & 2.87                    & -2.85                   & 7.69                      & 0.78                        & 1.32                         \\ \cline{2-7}
                         & UP06                        & 3.01                    & -2.99                   & 8.06                      & 0.58                        & 1.44                         \\ \cline{2-7} 
                         & UP09                        & 2.59                    & -2.59                   & 8.46                      & 0.90                        & 1.21                         \\ \cline{2-7} 
                         & UP11                        & 3.32                    & -3.32                   & 7.43                      & 0.97                        & 1.90                         \\ \cline{2-7} 
                         & UP16                        & 2.64                    & -2.64                   & 8.12                      & 0.91                        & 1.23                         \\ \cline{2-7} 
                         & UP18                        & 2.37                    & -2.37                   & 7.78                      & 1.17                        & 1.15                         \\ \cline{2-7} 
                         & UP20                        & 2.60                    & -2.60                   & 6.54                      & 0.81                        & 1.81                         \\ \cline{2-7} 
                         & UP22                        & 2.71                    & -2.71                   & 7.30                      & 0.55                        & 1.64                         \\ \cline{2-7} 
                         & UP25                        & 2.49                    & -2.47                   & 7.01                      & 0.63                        & 1.55                         \\ \cline{2-7} 
                         & UP36                        & 3.23                    & -3.23                   & 7.09                      & 0.54                        & 1.71                         \\ \cline{2-7} 
                         & UP39                        & 3.24                    & -3.22                   & 7.13                      & 0.63                        & 1.68                         \\ \cline{2-7} 
                         & UP40                        & 3.10                    & -3.10                   & 7.04                      & 0.98                        & 1.36                         \\ \cline{2-7} 
                         & UP41                        & 3.48                    & -3.48                   & 8.25                      & 0.56                        & 1.18                         \\ \cline{2-7} 
                         & UP42                        & 2.56                    & -2.51                   & 6.43                      & 0.94                        & 1.44                         \\ \cline{2-7} 
                         & UP43                        & 2.91                    & -2.91                   & 8.10                      & 0.94                        & 1.16                         \\ \cline{2-7} 
                         & X25                         & 2.84                    & -2.83                   & 7.90                      & 0.49                        & 1.45                         \\ \cline{2-7} 
                         & \textbf{Average}                    & \textbf{2.86}                    & \textbf{-2.84}                   & \textbf{7.53}                      & \textbf{0.70}                        & \textbf{1.42}                         \\ \hline
\end{tabularx}
\caption{Detection performance on the held-out SJO03 dataset and independent validation dataset from 16 UPFLOW stations.}
\label{table:evaluation_results}
\end{table}
\clearpage

\subsection{Seasonal and spatial patterns in call occurrence}

\begin{figure}[h!]
  \includegraphics[width=\textwidth, trim={0 0 0 1.3cm}, clip]{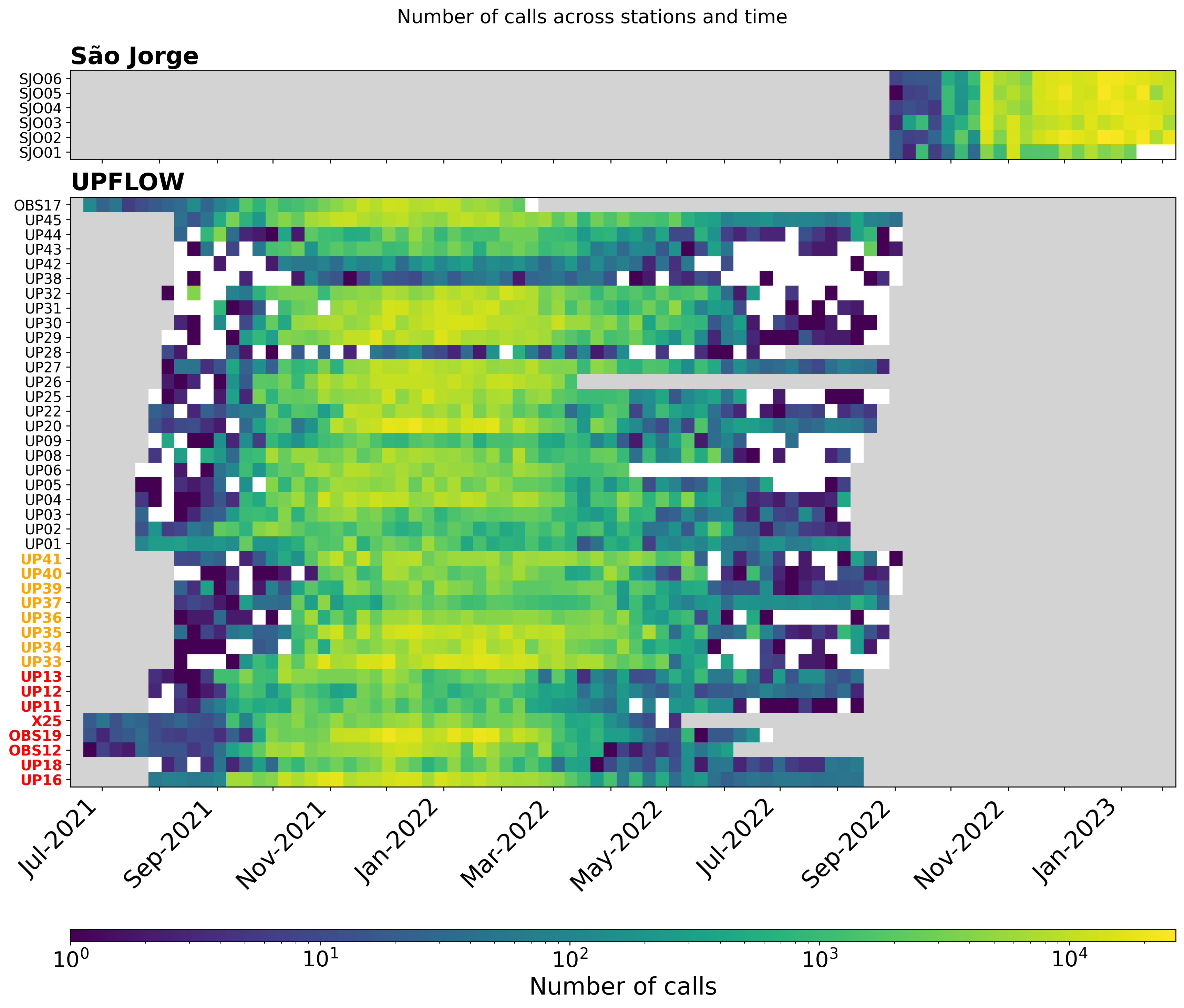}
  \caption{Weekly fin whale call counts across all 46 stations in this study. The heatmap shows the number of detected calls per station aggregated by week. Grey intervals correspond to periods where the instrument was not recording, while white intervals indicate weeks with no detections. Elevated call abundance is concentrated between October and February across most stations. North-western (located farther offshore in the central North Atlantic) and south-eastern -most stations (located closer to continental Portugal and Morocco) are highlighted in red and yellow, respectively.}
  \label{fig:call_count_heatbars}
\end{figure}

Weekly aggregation of detections reveals coherent seasonal patterns within both arrays (Fig. \ref{fig:call_count_heatbars}). Call activity increased noticeably from October and remained elevated through February across nearly all UPFLOW stations, with sparse or no detections outside this window. 


Detection rates also varied spatially.  North-western UPFLOW stations, located farther offshore in the central North Atlantic, recorded higher call counts earlier in the singing season relative to south-eastern stations located closer to the continental margins of Portugal and Morocco. This spatial contrast is visible in both the weekly heatmaps (Fig.~\ref{fig:call_count_heatbars}) and the aggregated time series by station group (Fig. \ref{fig:UPFLOW_compare_callcount_INI_bubble}b).

\subsection{INI distribution}

To confirm catalogue reliability for song timing analysis, detected INIs were compared with manual annotations at SJO03 across 14,280 matched pairs. The mean INI error was 0.80~s, the median error 0.045~s, and the MAE 2.56~s. Given that typical INI values in this region range between 10-20~s, this accuracy is sufficient to characterise inter-note structure. Representative singlet and doublet interval patterns reproduced by our catalogue are shown in Fig.~\ref{fig:Song_INI_comparisons}.

\begin{figure}[h]
\centering
\begin{subfigure}[b]{.475\textwidth}
    \includegraphics[width=\linewidth, trim={0 0 0 3cm}, clip]{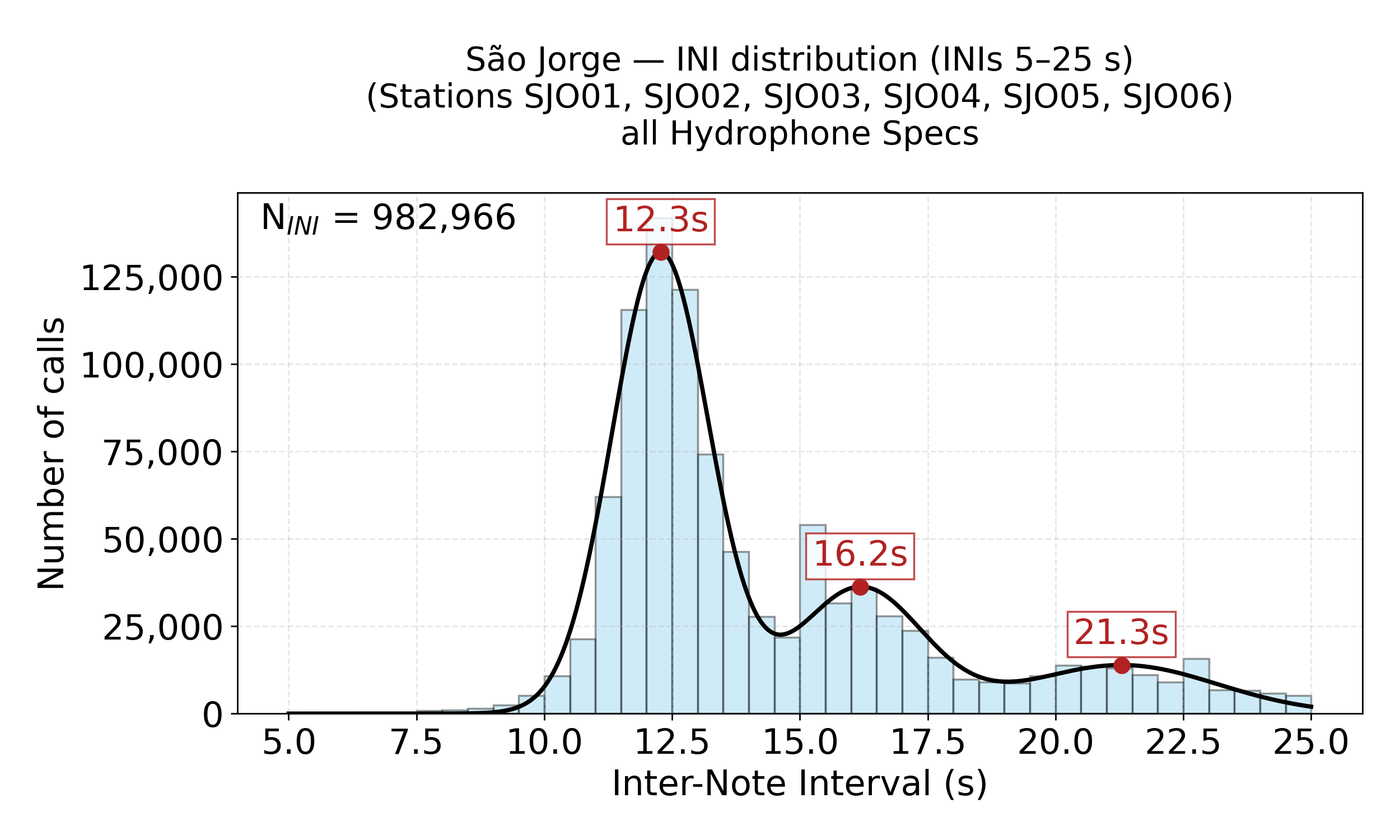}
        \put(-240,100){\large a)}             
\end{subfigure}
\hspace{0.5cm}
\begin{subfigure}[b]{.485\textwidth}
    \includegraphics[width=\linewidth,trim={1.3cm 0 0 6cm}, clip]{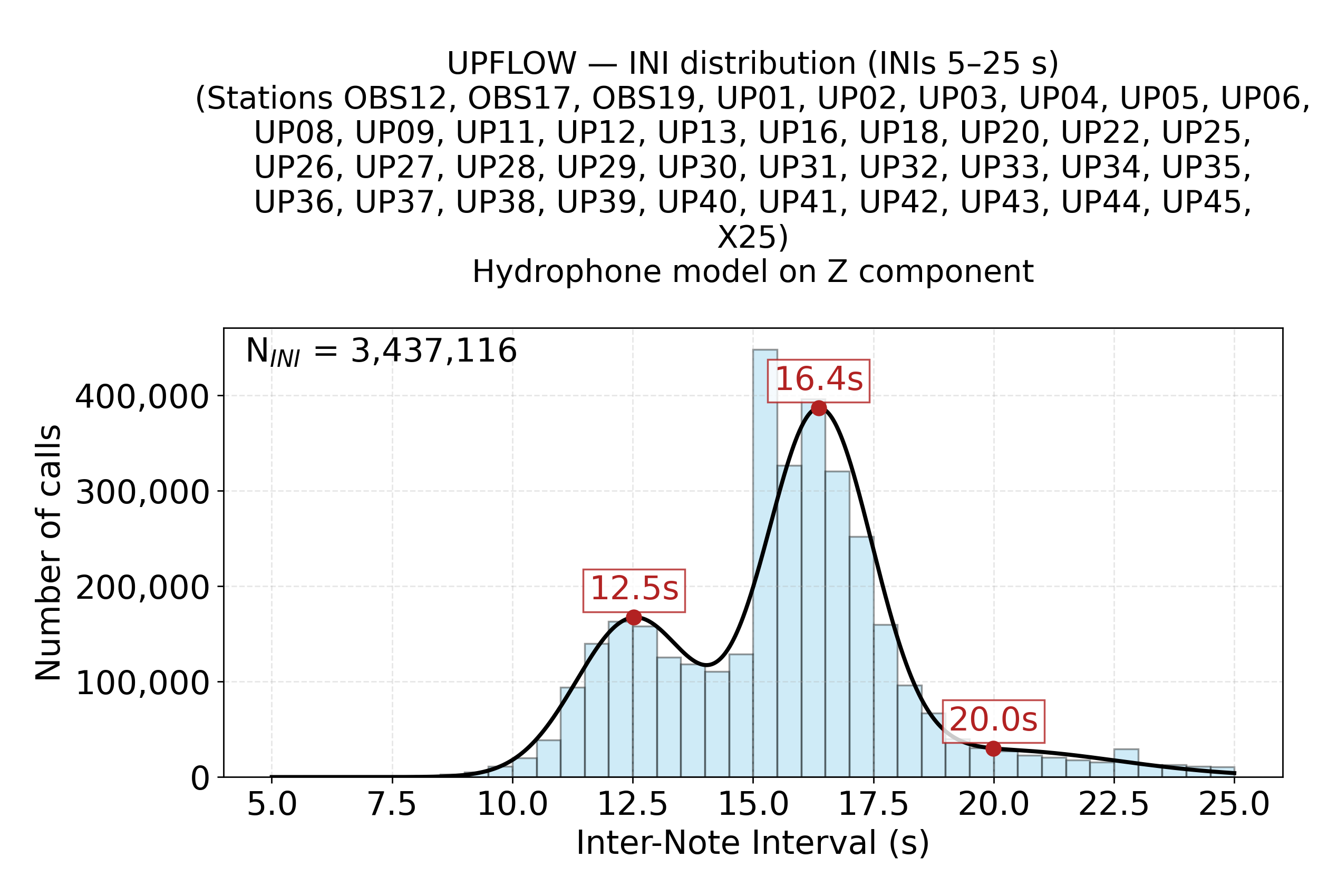}
        \put(-240,100){\large b)}             
\end{subfigure}
\caption{INI distributions of 20-Hz notes at the \textbf{a)} São Jorge array (6 stations, August 2022 to February 2023) and the \textbf{b)} UPFLOW array (40 stations, June 2021 to August 2022). Three distinct INI groups are observed in both arrays.
}
\label{fig:mean_INI_GMM_distributions}
\end{figure}

\begin{figure}[h!]
      \includegraphics[width=\textwidth, trim={0 0 0 2cm}, clip]{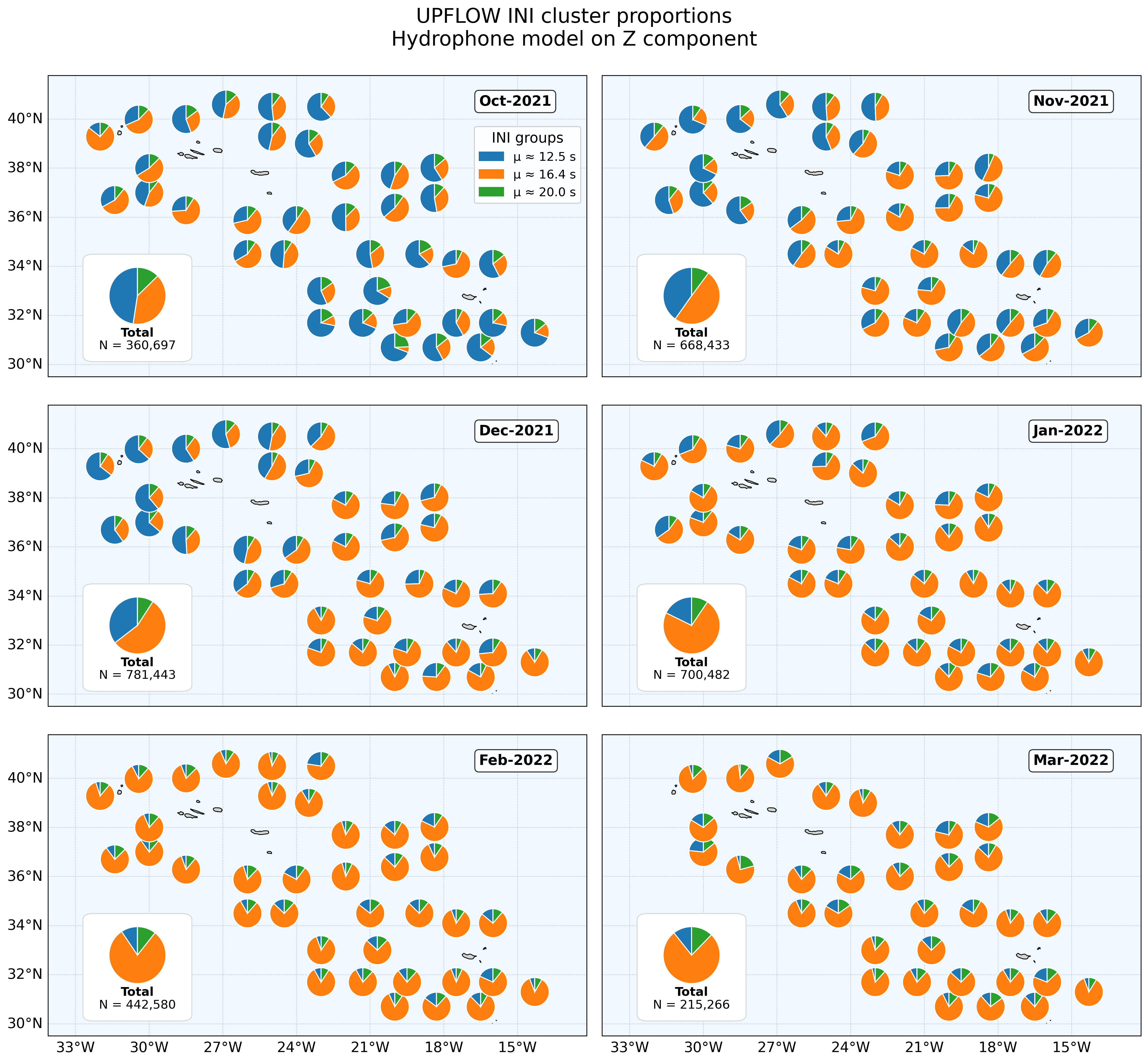}
      \caption{INI group composition of fin whale calls on the UPFLOW array during peak singing season. Detections are clustered into one of three INI groups ($\mu$). Stations with fewer than 1,000 INIs for a given month are excluded from the plot.
      }
      \label{fig:INI_cluster}
\end{figure}

Three distinct INI groups, centred at $\sim$12~s, 16~s, and 20~s, appeared at almost every station in both arrays  (Fig. \ref{fig:mean_INI_GMM_distributions}). The composition of these groups shifts coherently across the singing season in the UPFLOW array (Fig. \ref{fig:INI_cluster}).  In October the 12~s group was most common, particularly at southern stations, and northern stations came to overtake this dominance through November and December. From January onward the 16~s group dominated across all stations and persisted through February and March, while the 20~s group remained a small but persistent component throughout. 

\begin{figure}[h!]
    \begin{flushright}
      \includegraphics[width=17cm,trim={0 0 0 2cm}, clip]
      {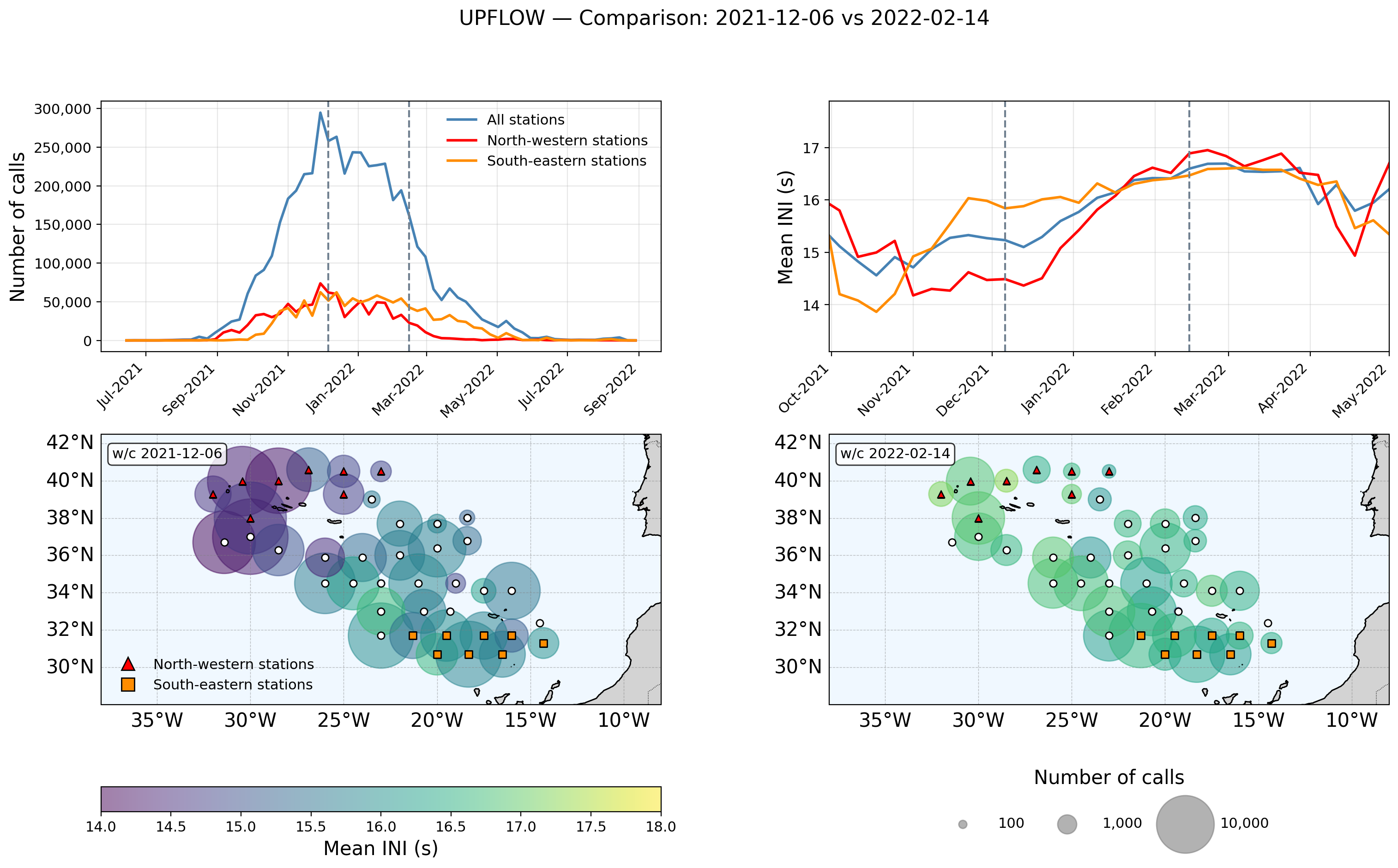}
      \put(-515,260){\large a)}   
      \put(-240,260){\large b)}   
      \put(-515,145){\large ci)}
      \put(-250,145){\large cii)}   
    
      \caption{\textbf{a)} Number of calls detected and \textbf{b)} mean INI over time in the UPFLOW array, aggregated by week. Only peak singing months are included in the latter. North-western and south-eastern -most stations are highlighted in red and yellow, respectively. Grey dashed vertical lines indicate the two weeks that contrast the start and end of the singing season. These are the weeks used in \textbf{c)}, which shows a bubble plot of the number of calls detected at each station (indicated by the size of the bubble), and the mean INI (indicated by the bubble's colour), in a single week beginning in \textbf{ci)} early December and \textbf{cii)} mid-February.
      }
      \label{fig:UPFLOW_compare_callcount_INI_bubble}
    \end{flushright}
\end{figure}

\begin{figure}[ht!]
  \includegraphics[width=\textwidth, trim={0 0 0 1.3cm}, clip]{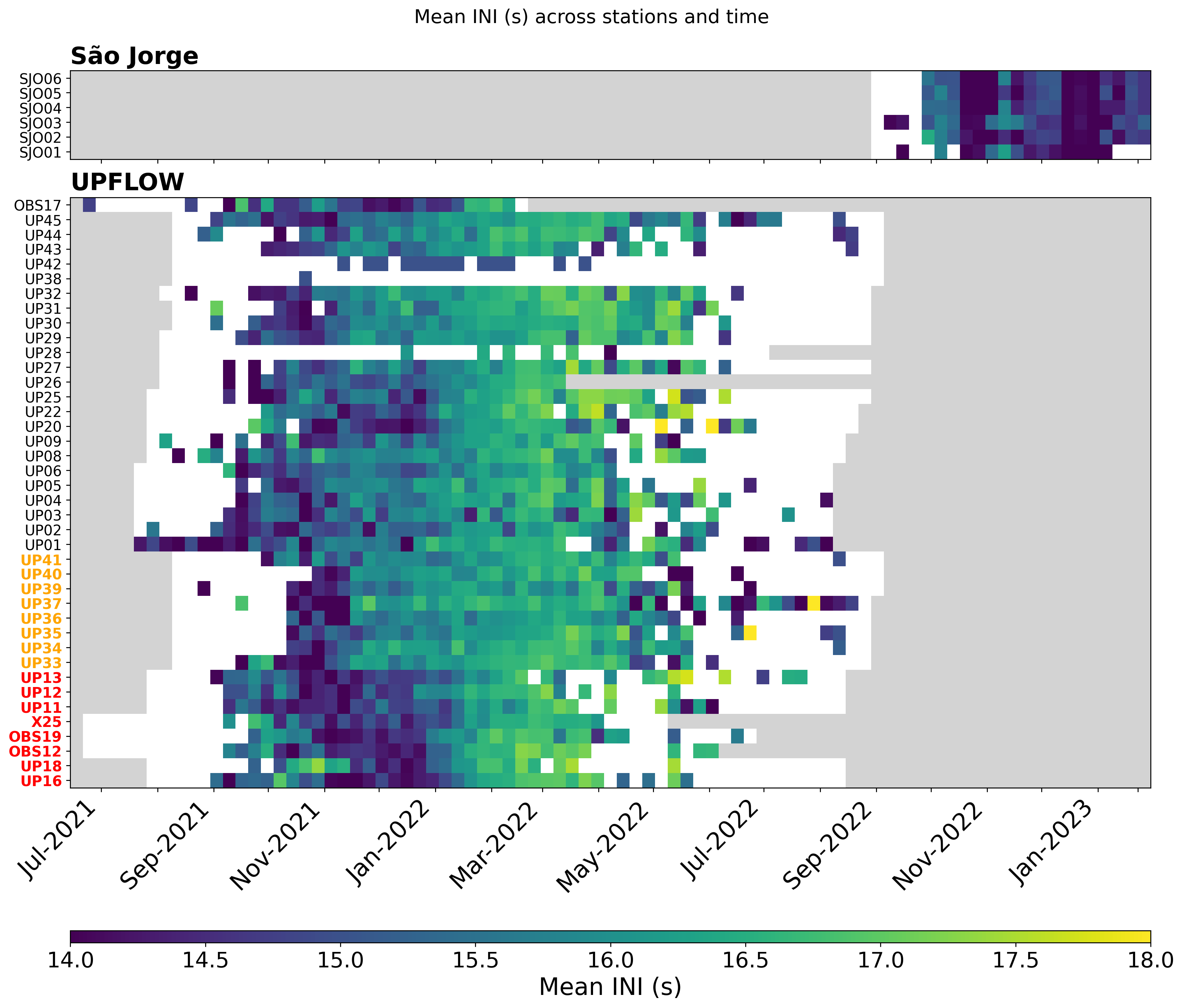}
  \caption{Heatmap of weekly mean INI values across all 46 stations in this study. Grey intervals correspond to periods where the instrument is not recording, while white intervals show insufficient INIs as only weeks with more than 100 valid INIs were considered. North-western and south-eastern -most stations are highlighted in red and yellow, respectively. }
  \label{fig:mean_INI_heatbars}
\end{figure}

Weekly summaries of mean INI further reveal spatial contrasts, with qualitative differences between north-western and south-eastern station groups during the peak calling period (Figs.~\ref{fig:UPFLOW_compare_callcount_INI_bubble} and \ref{fig:mean_INI_heatbars}).

\newpage

\section{Discussion} \label{Discussion}

\subsection{Detector performance}

Our segmentation-based architecture generates dense predictions in the time-frequency plane, enabling the direct estimation of onset time, duration, and frequency bounds for each detected call. This contrasts with classification-based detectors that report only event occurrence within a fixed time window \parencite[e.g.,][]{ruiz2025automated,edwards2026seismic,garcia2020comparing}, and enables downstream analyses from physically interpretable call parameters without requiring separate post-hoc measurement.  The suitability of pixel-level inference for bioacoustic detection is further supported by \textcite{jin2022semantic}, who applied a comparable segmentation approach to dolphin whistle extraction. 

A key feature of this study is that the model was trained on hydrophone data from the São Jorge array and subsequently applied to vertical-component seismometer recordings from the UPFLOW array. These sensor types record the same acoustic source through different coupling pathways, and differences in instrument response and local propagation conditions alter the appearance of calls in a spectrogram. The UPFLOW array also spans a much wider geographic area than São Jorge and samples a broad range of propagation environments, so the cross-sensor transfer was tested under substantially different recording conditions. Despite this, precision remained high across UPFLOW stations (>93\%), demonstrating that the learned spectral representation can meaningfully transfer across sensor types and recording environments. \textcite{dreo2025singing} similarly found a strong correlation between hydrophone and vertical seismometer channels for detecting baleen whale calls using OBS data, supporting the viability of using seismometers for PAM.


The low and variable recall across UPFLOW stations (9-53\%) reflects the diverse and variable recording conditions across the UPFLOW deployment rather than inconsistency in the learned call representation. The array spans a large geographic area with variable bathymetry, water depth, sediment properties, and oceanographic conditions, all of which influence how acoustic energy couples into the seafloor and propagates to a given instrument. Some environmental factors also vary seasonally, further contributing to site-specific and time-varying detectability. Stations with persistently elevated noise floors, such as UP28 and UP38, recorded anomalously low detection counts throughout their deployments (Table~\ref{table: station detection summary}), consistent with strong masking of fin whale vocalisations by broadband or narrowband noise (Figs.~\ref{fig:anomaly_UP28}-\ref{fig:anomaly_UP39}). Other sources of missed detections include low SNR, environmental and anthropogenic noise, and limited call separability during dense chorusing events (Fig.~\ref{fig: Spectrograms_masking_examples}), all of which are well-recognised challenges in automated PAM \parencite{gibb2019emerging}. These factors also contribute to the lower recall observed on UPFLOW relative to São Jorge. A more detailed characterisation of missed detections across recording conditions would be a valuable direction for future work.


\subsection{Spatiotemporal patterns in call occurrence and song structure}

The concentration of detected calls between October and February across the UPFLOW array aligns with known patterns of fin whale acoustic activity in the northeast Atlantic \parencite{romagosa2020baleen, nieukirk2004low} and with acoustic detections off the southwest of Portugal, where calling also peaks in autumn and winter \parencite{pereira2020fin}. Our results extend this seasonal signature across the broader Madeira-Azores-Canaries region, a largely open-ocean area where acoustic data on fin whales are sparse, and help fill an important geographical and temporal gap in North Atlantic fin whale monitoring. The high levels of singing activity observed throughout the area may indicate that fin whales are engaged in breeding-related behaviours, as fin whale songs are widely believed to function as reproductive displays \parencite{watkins198720, croll2002only}.

Calling activity first appeared at north-western stations earlier in the season before increasing at south-eastern stations closer to the continental margins, a pattern that could be consistent with a southward movement of singing whales from higher latitudes as the season progresses. Satellite tracking has demonstrated that fin whales tagged off Svalbard can reach southwestern Portugal and Morocco by late November and December \parencite{lydersen2020autumn}, and stable isotope evidence suggests that some fin whales passing through the Azores in spring may have wintered off the Iberian coast \parencite{silva2019stable}, indicating that north-to-south movements through the region do occur. We note, however, that this interpretation is speculative. Fin whales do not always follow predictable seasonal migratory patterns, and individual movement strategies in the Northeast Atlantic are known to be highly variable \parencite{silva2013north, lydersen2020autumn}. Furthermore, as previously discussed, spatial differences in detectability driven by variation in propagation conditions and noise levels may also independently influence the apparent calling activity. Call counts can also vary with INI, since shorter intervals produce higher instantaneous call rates, though this is likely a secondary effect relative to detectability differences.

The three persistent INI groups identified across both deployments, centred at  $\sim$12~s, 16~s, and 20~s (Fig.~\ref{fig:mean_INI_GMM_distributions}), are consistent with values reported across the wider North Atlantic \parencite[e.g.,][]{guazzo2024decade,pereira2020fin,romagosa2024fin}. Their composition shifted coherently across the singing season in the UPFLOW array (Fig. \ref{fig:INI_cluster}), with different INI groups dominant at different times of year across the array. Intra-seasonal variation in song INIs has been documented in the North Pacific, where INIs reset to similar values at the start of each season across widely separated monitoring sites \parencite{oleson2014synchronous}, and the drivers of such variation remain poorly understood. The scale and continuity of our catalogue, spanning 46 stations and approximately one year of continuous recording across the Azores-Madeira-Canaries region, provide a stronger empirical foundation for investigating these questions than has previously been available here. Rather than subsampling individual tracks, we processed all available recordings continuously, reducing susceptibility to song sampling biases and enabling INI variation to be characterised at a geographic and temporal resolution not previously achieved in this region. We therefore view the catalogue as a valuable resource for future work aimed at understanding the mechanisms driving song variation in fin whales. A more complete characterisation of singing patterns would also require analysis of other call types, including the less abundant \textit{'backbeat'}, and other INI pairings \parencite{helble2020fin}, which our current catalogue does not capture and which represent a natural direction for future work.


\subsection{Broader implications and future directions}

Large-scale PAM of baleen whales has historically been constrained by the cost and spatial coverage of dedicated hydrophone infrastructure. This study demonstrates that repurposing existing seismo-acoustic datasets can substantially reduce this barrier. OBS deployments are ongoing across the world's ocean basins, and the low-frequency band they record is shared by many baleen whale species of high conservation concern. Applied opportunistically to these archives, automated detection frameworks like that presented here could expand the spatial and temporal coverage of PAM at comparatively low cost, at scales that better match the ecological range of highly mobile species than is achievable through traditional point-based sampling.

The catalogue produced here is released with this paper (see Data availability) and opens several concrete avenues for ecological research. Long-duration catalogues spanning multiple deployments could enable tracking of inter- and intra-annual changes in song structure, building on the long-term INI trends documented by \textcite{romagosa2024fin} and \textcite{guazzo2024decade}, and improving our understanding of the factors driving and constraining fin whale song variation. Spatially and temporally resolved call rates can in principle be used to estimate fin whale abundance across the North Atlantic, supplementing visual survey data during periods and in areas where surveys are not feasible \parencite{marques2013estimating}. The UPFLOW array in particular covers large areas of remote waters where fin whale distribution remains poorly documented, and the catalogue begins to fill this important gap. At a management level, such data could also inform marine spatial planning, including identification of areas and periods of elevated acoustic activity relevant to shipping route design or assessment of anthropogenic noise impacts on singing behaviour \parencite[e.g.,][]{castellote2012acoustic,edwards2026seismic}.

More broadly, call detection is often the first step in a PAM workflow, enabling downstream analyses from abundance estimation to assessing the effects of environmental and anthropogenic covariates on distribution and behaviour. The same detection framework could be extended to other migratory species whose calls occupy similar frequency ranges, including blue, sei, and humpback whales, substantially expanding the ecological value of existing geophysical archives.

\section{Conclusion} \label{Conclusion}

OBSs represent a largely untapped resource for marine ecological research. By repurposing two deployments in the eastern North Atlantic, we produced the largest catalogue of fin whale 20-Hz notes to date, demonstrating that automated deep learning can unlock the ecological value of existing seismo-acoustic archives at basin scale and low cost. The cross-sensor performance of our detector shows that this approach can be applied to varying sites, broadening its potential applicability across other deployments. Our catalogue reveals coherent seasonal calling patterns and complex intra-seasonal INI dynamics across a largely unmonitored region, illustrating the kind of population-level insight that only becomes accessible through continuous, large-scale acoustic monitoring. Understanding what drives these behavioural patterns will require the kind of spatially dense, long-duration datasets that OBS recordings can uniquely provide. Frameworks like that presented here, extended to other low-frequency species and applied to ongoing deployments worldwide, could substantially advance our ability to monitor, understand, and protect large baleen whale populations at the scales their ecology demands.


\printbibliography

\begin{acknowledgements}
Jocelyn Japnanto was supported by the UKRI Engineering and Physical Sciences Research Council (EPSRC) [grant number EP/T517793/1 and EP/W524335/1]. The UPFLOW project is funded by the ERC Horizon Fund (grant agreement number 101001601, PI A. M. G. Ferreira) and the São Jorge OBS data were collected thanks to a Natural Environment Research Council (NERC) Urgency Grant (NE/X006298/1 to A.M.G.F. and co-Is Ricardo Ramalho and Neil Mitchell). We are grateful to the UK Ocean Bottom Instrument Consortium (OBIC) (https://obs.ac.uk) for providing the instrumentation and installation services around São Jorge Island. The Portuguese Navy (Marinha Portuguesa) is acknowledged for providing critical support during the deployment and recovery of the São Jorge OBS network; we particularly thank the captains and crews of the NRPs António Enes and Sines. The Azores Government, through its Fundo Regional para a Ciência, is also acknowledged for its financial support to harbour operations during the deployment and recovery of the São Jorge OBS network. We thank Carlos Corela, Gonçalo Henriques, Paula Lourinho, Filipe Porteiro and Octávio Melo from OKEANOS, University of the Azores, for crucial help and logistics support with the São Jorge OBS deployment. We are grateful to all the institutions that contributed instruments to the UPFLOW experiment: 32 OBSs were rented from the DEPAS international pool of instruments maintained by the Alfred Wegener Institute (Bremerhaven), Germany (https://www.awi.de/en/science/geosciences/geophysics/methods-and-tools/ocean-bottom-seismometer/depas.html), while additional instruments were borrowed from other institutions, 7 from DIAS’ iMARL OBS pool (https://geohub.dias.ie/imarl/), 4 from IDL, 3 from ROA-UCM, and 4 from GEOMAR. We are also grateful to IPMA (co-beneficiary of the UPFLOW project), who made the research vessel NI Mário (https://marioruivo.ipma.pt/en/home-page/) Ruivo available for the UPFLOW experiment.

\end{acknowledgements}

\section*{Data availability statement}
\textit{The codebase is available on Github: \url{https://anonymous.4open.science/r/WhaleSeg-2B13/} . The full detection catalogue can be found in Zenodo: \url{https://doi.org/10.5281/zenodo.21065182}}

\section*{Conflict of interest}  
      
The authors declare no conflict of interest.

\newpage

\setcounter{section}{0}
\setcounter{figure}{0}
\setcounter{table}{0}
\setcounter{equation}{0}

\setcounter{page}{1}
\resetlinenumber[1]

\renewcommand{\thesection}{S\arabic{section}}
\renewcommand{\thefigure}{S\arabic{figure}}
\renewcommand{\thetable}{S\arabic{table}}
\renewcommand{\theequation}{S\arabic{equation}}

\section{Supplementary Material}

\subsection{Station recording characteristics}\label{SI: station table}

{\renewcommand{\arraystretch}{0.98}
\begin{table}[h!]
\begin{tabular}{|c|c|c|c|c|c|c|}
\hline
\textbf{Network}           & \textbf{Station} & \textbf{\makecell{Deployment\\start}} & \textbf{\makecell{Recovery\\date}} & \multicolumn{1}{c|}{\textbf{Sensor used}}       & \textbf{\makecell{Sampling\\rate (Hz)}} & \textbf{\makecell{Station\\depth (km)}} \\ \hline
\multirow{6}{*}{São Jorge} & SJO01            & 26/08/2022                & 30/01/2023             & \multirow{6}{*}{\makecell{HighTech\\HTI-96-Min\\hydrophone}} & 250                         & 1.24                        \\ \cline{2-4} \cline{6-7} 
                           & SJO02            & 26/08/2022                & 30/01/2023             &                                                 & 250                         & 1.36                        \\ \cline{2-4} \cline{6-7} 
                           & SJO03            & 26/08/2022                & 29/01/2023             &                                                 & 250                         & 1.24                        \\ \cline{2-4} \cline{6-7} 
                           & SJO04            & 26/08/2022                & 29/01/2023             &                                                 & 250                         & 1.16                        \\ \cline{2-4} \cline{6-7} 
                           & SJO05            & 26/08/2022                & 29/01/2023             &                                                 & 250                         & 1.16                        \\ \cline{2-4} \cline{6-7} 
                           & SJO06            & 26/08/2022                & 29/01/2023             &                                                 & 250                         & 1.13                        \\ \hline
\multirow{40}{*}{UPFLOW}   & OBS12            & 19/06/2021                & 02/06/2022             & \multirow{40}{*}{\makecell{Trillium\\Compact}}              & 100                         & 2.10                        \\ \cline{2-4} \cline{6-7} 
                           & OBS17            & 21/06/2021                & 01/01/2022             &                                                 & 100                         & 2.65                        \\ \cline{2-4} \cline{6-7} 
                           & OBS19            & 19/06/2021                & 20/06/2022             &                                                 & 100                         & 1.92                        \\ \cline{2-4} \cline{6-7} 
                           & UP01             & 17/07/2021                & 04/08/2022             &                                                 & 250                         & 5.46                        \\ \cline{2-4} \cline{6-7} 
                           & UP02             & 18/07/2021                & 05/08/2022             &                                                 & 250                         & 4.84                        \\ \cline{2-4} \cline{6-7} 
                           & UP03             & 18/07/2021                & 05/08/2022             &                                                 & 250                         & 4.55                        \\ \cline{2-4} \cline{6-7} 
                           & UP04             & 18/07/2021                & 06/08/2022             &                                                 & 250                         & 5.27                        \\ \cline{2-4} \cline{6-7} 
                           & UP05             & 19/07/2021                & 06/08/2022             &                                                 & 100                         & 4.22                        \\ \cline{2-4} \cline{6-7} 
                           & UP06             & 19/07/2021                & 06/08/2022             &                                                 & 100                         & 4.44                        \\ \cline{2-4} \cline{6-7} 
                           & UP08             & 20/07/2021                & 08/08/2022             &                                                 & 100                         & 4.34                        \\ \cline{2-4} \cline{6-7} 
                           & UP09             & 21/07/2021                & 08/08/2022             &                                                 & 100                         & 3.87                        \\ \cline{2-4} \cline{6-7} 
                           & UP11             & 21/07/2021                & 09/08/2022             &                                                 & 100                         & 4.22                        \\ \cline{2-4} \cline{6-7} 
                           & UP12             & 22/07/2021                & 10/08/2022             &                                                 & 100                         & 3.45                        \\ \cline{2-4} \cline{6-7} 
                           & UP13             & 22/07/2021                & 12/08/2022             &                                                 & 100                         & 3.30                        \\ \cline{2-4} \cline{6-7} 
                           & UP16             & 24/07/2021                & 13/08/2022             &                                                 & 250                         & 1.44                        \\ \cline{2-4} \cline{6-7} 
                           & UP18             & 24/07/2021                & 14/08/2022             &                                                 & 100                         & 1.95                        \\ \cline{2-4} \cline{6-7} 
                           & UP20             & 25/07/2021                & 17/08/2022             &                                                 & 250                         & 3.32                        \\ \cline{2-4} \cline{6-7} 
                           & UP22             & 26/07/2021                & 18/08/2022             &                                                 & 100                         & 3.31                        \\ \cline{2-4} \cline{6-7} 
                           & UP25             & 26/07/2021                & 22/08/2022             &                                                 & 100                         & 4.14                        \\ \cline{2-4} \cline{6-7} 
                           & UP26             & 27/07/2021                & 10/03/2022             &                                                 & 100                         & 5.13                        \\ \cline{2-4} \cline{6-7} 
                           & UP27             & 27/07/2021                & 23/08/2022             &                                                 & 250                         & 4.91                        \\ \cline{2-4} \cline{6-7} 
                           & UP28             & 27/07/2021                & 28/06/2022             &                                                 & 100                         & 4.98                        \\ \cline{2-4} \cline{6-7} 
                           & UP29             & 28/07/2021                & 24/08/2022             &                                                 & 100                         & 5.26                        \\ \cline{2-4} \cline{6-7} 
                           & UP30             & 03/08/2021                & 25/08/2022             &                                                 & 250                         & 5.03                        \\ \cline{2-4} \cline{6-7} 
                           & UP31             & 03/08/2021                & 25/08/2022             &                                                 & 100                         & 5.19                        \\ \cline{2-4} \cline{6-7} 
                           & UP32             & 02/08/2021                & 25/08/2022             &                                                 & 100                         & 5.33                        \\ \cline{2-4} \cline{6-7} 
                           & UP33             & 04/08/2021                & 26/08/2022             &                                                 & 100                         & 4.95                        \\ \cline{2-4} \cline{6-7} 
                           & UP34             & 04/08/2021                & 26/08/2022             &                                                 & 100                         & 4.80                        \\ \cline{2-4} \cline{6-7} 
                           & UP35             & 05/08/2021                & 27/08/2022             &                                                 & 250                         & 4.60                        \\ \cline{2-4} \cline{6-7} 
                           & UP36             & 05/08/2021                & 28/08/2022             &                                                 & 100                         & 4.24                        \\ \cline{2-4} \cline{6-7} 
                           & UP37             & 06/08/2021                & 28/08/2022             &                                                 & 250                         & 3.65                        \\ \cline{2-4} \cline{6-7} 
                           & UP38             & 06/08/2021                & 29/08/2022             &                                                 & 100                         & 4.39                        \\ \cline{2-4} \cline{6-7} 
                           & UP39             & 06/08/2021                & 29/08/2022             &                                                 & 100                         & 4.44                        \\ \cline{2-4} \cline{6-7} 
                           & UP40             & 07/08/2021                & 29/08/2022             &                                                 & 100                         & 4.51                        \\ \cline{2-4} \cline{6-7} 
                           & UP41             & 07/08/2021                & 30/08/2022             &                                                 & 100                         & 4.38                        \\ \cline{2-4} \cline{6-7} 
                           & UP42             & 07/08/2021                & 30/08/2022             &                                                 & 100                         & 3.86                        \\ \cline{2-4} \cline{6-7} 
                           & UP43             & 08/08/2021                & 31/08/2022             &                                                 & 100                         & 5.08                        \\ \cline{2-4} \cline{6-7} 
                           & UP44             & 09/08/2021                & 01/09/2022             &                                                 & 100                         & 4.27                        \\ \cline{2-4} \cline{6-7} 
                           & UP45             & 09/08/2021                & 01/09/2022             &                                                 & 250                         & 4.02                        \\ \cline{2-4} \cline{6-7} 
                           & X25              & 20/06/2021                & 03/05/2022             &                                                 & 100                         & 2.52                        \\ \hline
\end{tabular}
\caption{Recording characteristics of each station from the São Jorge and UPFLOW arrays used in this study.}
\label{SI: station detection summary}
\end{table}
}


\subsection{Training data samples}\label{SI: Metrics and Notation}

\begin{figure}[h!]
  \includegraphics[width=\textwidth, trim={0 0cm 0 1cm}, clip]{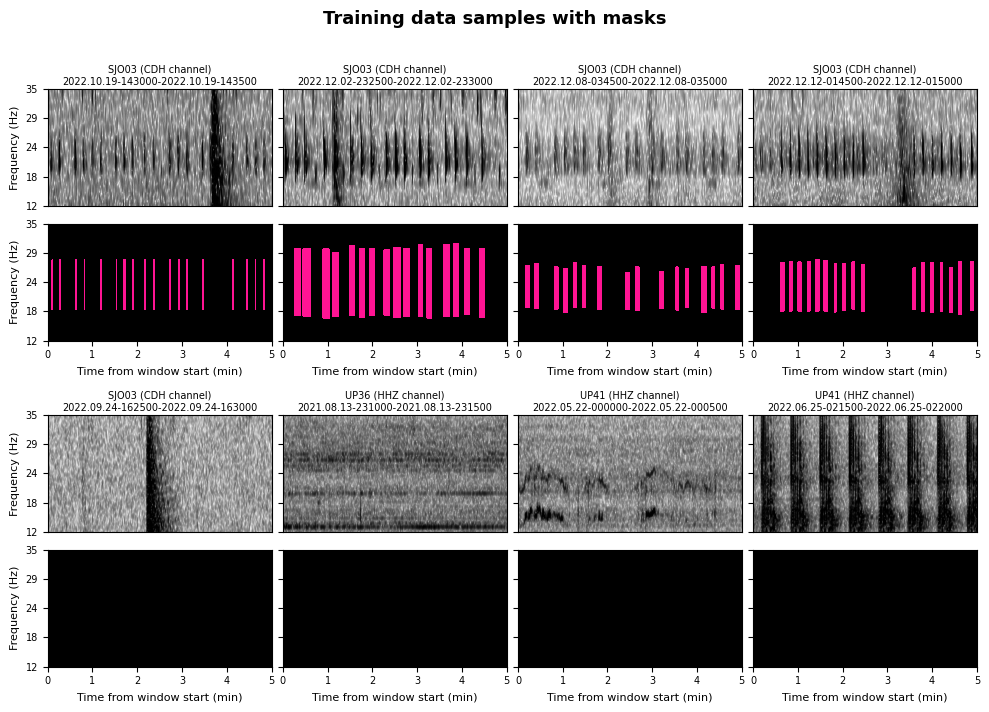}
  \caption{Samples from the training set. \textbf{Top:} Samples from station SJO03 with annotated calls (pink). \textbf{Bottom:} Background samples from multiple stations, showing various sources of noise.}
  \label{fig:training_samples}
\end{figure}


\subsection{Metrics and Notation}\label{SI: Metrics and Notation}

\subsubsection*{Filtering criteria}

These thresholds were derived from the distribution of call parameters in the annotation catalogue and are intended to remove detections that are greatly inconsistent with the known spectral and temporal properties of fin whale 20-Hz notes.

\begin{equation}\label{eq: filtering criteria}
\begin{split}
     21.5~\text{Hz} \geq f_{low}\\
     f_{high}\geq22.5~\text{Hz}\\
     \mathrm{BW} \geq 4.0~\mathrm{Hz}\\
     \mathrm{D} \geq  2.0~\mathrm{s}\\
\end{split}
\end{equation}


\subsubsection*{Bounding box representation}\label{subsection: Bounding box representation}

Each detection and annotation is represented as a time-frequency bounding box
\begin{equation}\label{eq: bounding box}
     B = (t_{\mathrm{start}}, t_{\mathrm{end}}, f_{\mathrm{low}}, f_{\mathrm{high}})
\end{equation}
where $t$ denotes time and $f$ denotes frequency. The parameters $t_{\mathrm{start}}$ and $t_{\mathrm{end}}$ correspond to the temporal onset and end of a call, while $f_{\mathrm{low}}$ and $f_{\mathrm{high}}$ denote the lower and upper frequency bounds of the call in the spectrogram.

For a given bounding box $B$, the call duration $D$ and bandwidth $BW$ are defined as
\begin{equation}
D = t_{\mathrm{end}} - t_{\mathrm{start}},
\end{equation}

\begin{equation}
\mathrm{BW} = f_{\mathrm{high}} - f_{\mathrm{low}}.
\end{equation}

Bounding boxes predicted by the detector are denoted $B_d$, while manual annotations are denoted $B_a$.


\subsubsection*{Matching procedure}\label{subsubsection: Matching procedure}

Predicted detections were matched to annotations based on the Intersection over Union (IoU) between their bounding boxes in time-frequency space.

For a detection box $B_d$ and an annotation box $B_a$, temporal and spectral overlap are computed as

\begin{equation}\label{eq:time_freq_intersection}
\begin{split}
\Delta t_{\cap} =
\max\left(0,\;
\min(t_d^{\mathrm{end}}, t_a^{\mathrm{end}})
-
\max(t_d^{\mathrm{start}}, t_a^{\mathrm{start}})
\right)
\\
\Delta f_{\cap} =
\max\left(0,\;
\min(f_d^{\mathrm{high}}, f_a^{\mathrm{high}})
-
\max(f_d^{\mathrm{low}}, f_a^{\mathrm{low}})
\right)
\end{split}
\end{equation}

The intersection area is

\begin{equation}\label{eq:intersection_area}
A_{\cap} = \Delta t_{\cap} \cdot \Delta f_{\cap}
\end{equation}

The areas of the detection and annotation boxes are

\begin{equation}\label{eq:det_ann_areas}
\begin{split}
A_d = (t_d^{\mathrm{end}} - t_d^{\mathrm{start}})
      (f_d^{\mathrm{high}} - f_d^{\mathrm{low}})
\\
A_a = (t_a^{\mathrm{end}} - t_a^{\mathrm{start}})
      (f_a^{\mathrm{high}} - f_a^{\mathrm{low}})
\end{split}
\end{equation}

The union area is

\begin{equation}\label{eq:union_area}
A_{\cup} = A_d + A_a - A_{\cap}
\end{equation}

and the IoU is therefore defined as

\begin{equation}\label{eq:iou}
    \mathrm{IoU}(B_d, B_a) =
    \frac{A_{\cap}}{A_{\cup}}
\end{equation}

A greedy one-to-one matching procedure \parencite{padilla2021comparative} was used, i.e., each $B_d$ is matched to one $B_a$, at most, and vice versa. For each detection, the IoU was computed with all annotations within the same 5-minute window. The annotation with the highest IoU was selected. A detection was considered a true positive (TP) if $\mathrm{IoU} > 0.15$. Unmatched detections were counted as false positives (FP), and unmatched annotations as false negatives (FN).

\subsubsection*{Detection metrics}\label{subsubsection: Detection metrics}
Performance was summarised using standard detection metrics:

\begin{itemize}

\item Precision: proportion of predicted detections that correspond to real calls

\begin{equation}\label{eq: Precision}
\mathrm{Precision} =
\frac{\mathrm{TP}}{\mathrm{TP + FP}}
\end{equation}

\item Recall: proportion of annotated calls successfully detected

\begin{equation}\label{eq: Recall}
\mathrm{Recall} =
\frac{\mathrm{TP}}{\mathrm{TP + FN}}
\end{equation}

\item F1-score: harmonic mean of precision and recall

\begin{equation}\label{eq: F1-score}
\mathrm{F1} =
\frac{2 \times \mathrm{Precision} \times \mathrm{Recall}}
{\mathrm{Precision} + \mathrm{Recall}}
=
\frac{2\,\mathrm{TP}}
{2\,\mathrm{TP} + \mathrm{FP} + \mathrm{FN}}
\end{equation}

\item Mean IoU across all matched detection-annotation pairs

\begin{equation}\label{eq: mean IoU}
\mathrm{mIoU} =
\frac{1}{N_{\mathrm{match}}}
\sum_{i=1}^{N_{\mathrm{match}}}
\mathrm{IoU}_i
\end{equation}

\end{itemize}


\subsubsection*{Temporal and spectral accuracy}

For each matched detection-annotation pair, localisation errors were also evaluated. Let $i$ index the $N_{\mathrm{match}}$ pairs, and let subscripts $d$ and $a$ denote detection and annotation, respectively.

The onset error measures the difference between detected and annotated start times:

\begin{equation}\label{eq:onset_error}
e^{\mathrm{onset}}_i =
t^{\mathrm{start}}_{d,i}
-
t^{\mathrm{start}}_{a,i}
\end{equation}

The mean absolute error and mean bias are

\begin{equation}\label{eq:onset_mae}
\mathrm{MAE}_{\mathrm{onset}} =
\frac{1}{N_{\mathrm{match}}}
\sum_{i=1}^{N_{\mathrm{match}}}
\left| e^{\mathrm{onset}}_i \right|
\end{equation}

\begin{equation}\label{eq:onset_bias}
\mathrm{Bias}_{\mathrm{onset}} =
\frac{1}{N_{\mathrm{match}}}
\sum_{i=1}^{N_{\mathrm{match}}}
e^{\mathrm{onset}}_i
\end{equation}

The duration error compares detected and annotated call durations:

\begin{equation}\label{eq:duration_error}
e^{\mathrm{dur}}_i =
D_{d,i} - D_{a,i}
\end{equation}

The mean absolute duration error is

\begin{equation}\label{eq:duration_mae}
    \mathrm{MAE}_{\mathrm{D}} =
    \frac{1}{N_{\mathrm{match}}}
    \sum_{i=1}^{N_{\mathrm{match}}}
    \left| e^{\mathrm{dur}}_i \right|
\end{equation}

Frequency localisation errors were also similarly computed for the lower and upper frequency bounds:

\begin{equation}\label{eq:f_low_mae}
    \mathrm{MAE}_{f_{\mathrm{low}}} =
    \frac{1}{N_{\mathrm{match}}}
    \sum_{i=1}^{N_{\mathrm{match}}}
    \left| e^{f_{\mathrm{low}}}_i \right|
\end{equation}

\begin{equation}\label{eq:fhigh_mae}
    \mathrm{MAE}_{f_{\mathrm{high}}} =
    \frac{1}{N_{\mathrm{match}}}
    \sum_{i=1}^{N_{\mathrm{match}}}
    \left| e^{f_{\mathrm{high}}}_i \right|
\end{equation}


\clearpage
\subsection{Signal-to-noise (SNR) analysis}\label{SI: SNR}
\subsubsection*{Call SNR estimation}

SNRs were computed for each detection using a spectrogram-based approach applied to short waveform segments centred on the detection time. For each event, a context window extending $\pm$15~s around the detection was extracted and instrument response removed to obtain ground velocity. A power spectrogram $S(f,t)$ (in linear units) was then computed using a short-time Fourier transform with a window length of 0.8 s and 95\% overlap. We defined our frequency band of interest $[f_{\min}, f_{\max}]$ to 18–26~Hz, corresponding to the dominant energy of the calls.

\begin{figure}[h!]
  \includegraphics[width=\textwidth, trim={0 0 0 0.7cm}, clip]{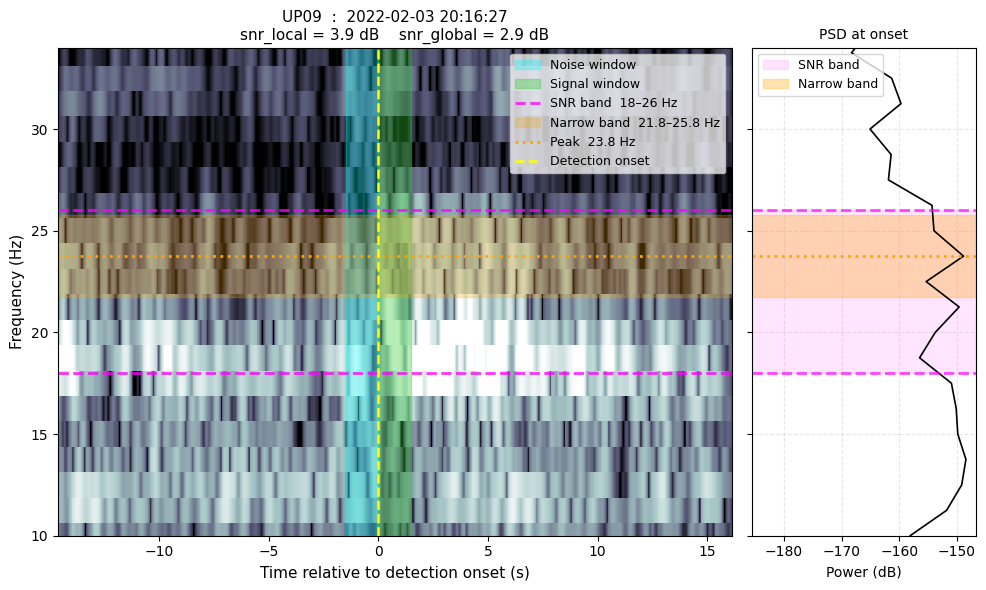}
  \caption{Example of spectrogram-based SNR estimation for a single detection from station UP09 (03-Feb-2022, 20:16:27). \textbf{Left}: Power spectrogram centred on the detection (yellow vertical line). SNR is computed within a predefined frequency band $[f_{\min}, f_{\max}]$ (magenta lines). The signal window (green) is defined as a duration $ T = 1.5~s$ immediately following the detection onset, while the noise window (cyan) is the preceding interval of equal duration. A narrow frequency band (orange) is constructed symmetrically about the dominant frequency (orange dotted line) at the onset time. The signal level $S_{\text{sig}}$ is taken as the median power within this narrow band over the signal window, whereas the local and global noise level $S_{\text{noise,local}}$ and $S_{\text{noise,global}}$ is the median power across the full frequency band within the noise window, and context window, respectively. \textbf{Right}: The power spectrum at the onset time and includes the peak frequency and the corresponding narrow-band selection used for SNR estimation.}
  \label{fig:SNR_calculation_example}
\end{figure}

The signal level was defined from a short time window of duration $T$ (1.5~s) beginning at the detection onset. The dominant frequency $f_{peak}$ was identified as the frequency bin with maximum power at the onset time. A narrow frequency band ($f_{peak} - \Delta f \;, \; f_{peak} + \Delta f $) was then constructed around this peak, where $\Delta f$ is a fixed bandwidth. These bands are visualised in Fig.~S\ref{fig:SNR_calculation_example}. The signal level was computed as the median spectrogram power within this time–frequency region:

\begin{equation}\label{eq: S_signal}
    S_{\text{sig}} = \mathrm{median}( S(f,t) \;\; | \;\; f \in [f_{peak} - \Delta f, f_{peak} + \Delta f] \;, \; t \in [t_0, t_0 + T])
\end{equation}


We define the local noise level from the time window of duration $T$ immediately preceding the detection, using the full analysis band:

\begin{equation}\label{eq: S_local_noise}
    S_{\text{noise,local}} = \mathrm{median}( S(f,t) \;\; | \;\; f \in [f_{\min}, f_{\max}] \;, \;  \in [t_0 - T, t_0]).
\end{equation}

and a global (ambient) noise level as the median spectrogram power over the entire context window:

\begin{equation}\label{eq: S_global_noise}
    S_{\text{noise,global}} = \mathrm{median}( S(f,t) \;\; | \;\; f \in [f_{\min}, f_{\max}] \;, \; t \in \text{context window}).
\end{equation}

SNRs were then expressed in decibels (dB) as:
\begin{equation}\label{eq: SNR_local}
    \mathrm{SNR}_{\text{local}} = 10 \log_{10} \left( \frac{S_{\text{sig}}}{S_{\text{noise,local}}} \right),
\end{equation}

\begin{equation}\label{eq: SNR_global}
    \mathrm{SNR}_{\text{global}} = 10 \log_{10} \left( \frac{S_{\text{sig}}}{S_{\text{noise,global}}} \right).
\end{equation}

\begin{figure}[h!]
  \includegraphics[width=\textwidth, trim={0 3cm 0 10.5cm}, clip]{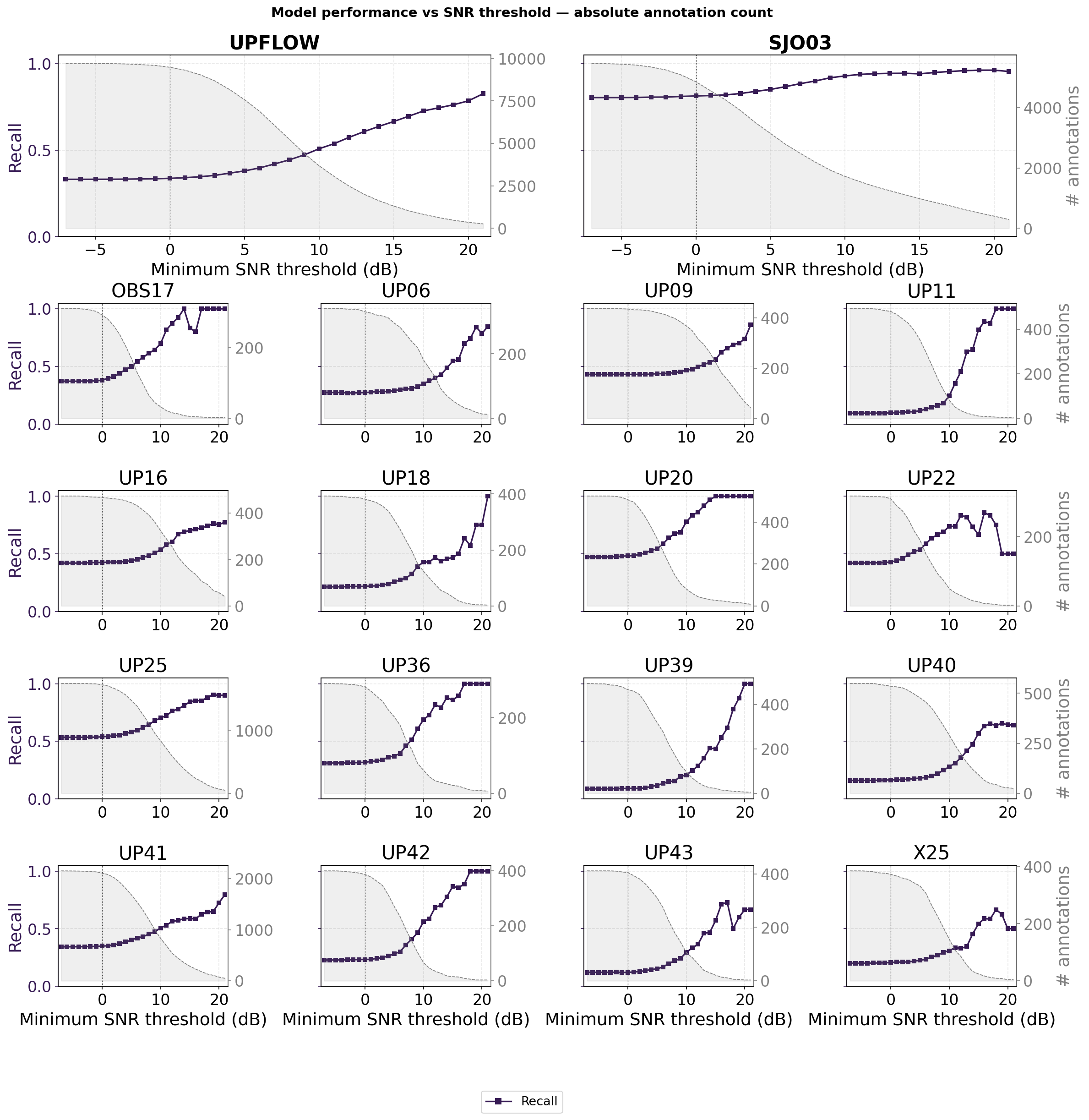}
  \caption{Recall as a function of minimum SNR threshold applied to ground-truth annotations, with results broken down at each annotated UPFLOW station.}
  \label{fig:SNR_recall_by_station}
\end{figure}

Fig.~S\ref{fig:SNR_recall_by_station} shows how the recall at each annotated UPFLOW station changes with varying SNR thresholds.

\clearpage
\subsection{Detections}\label{SI: results}

\subsubsection*{INI patterns in recorded songs}

\begin{figure}[h!]
\centering
\begin{subfigure}{\textwidth}
    \centering
    \begin{overpic}[trim=0 1.2cm 0 2cm,clip,width=.96\linewidth]{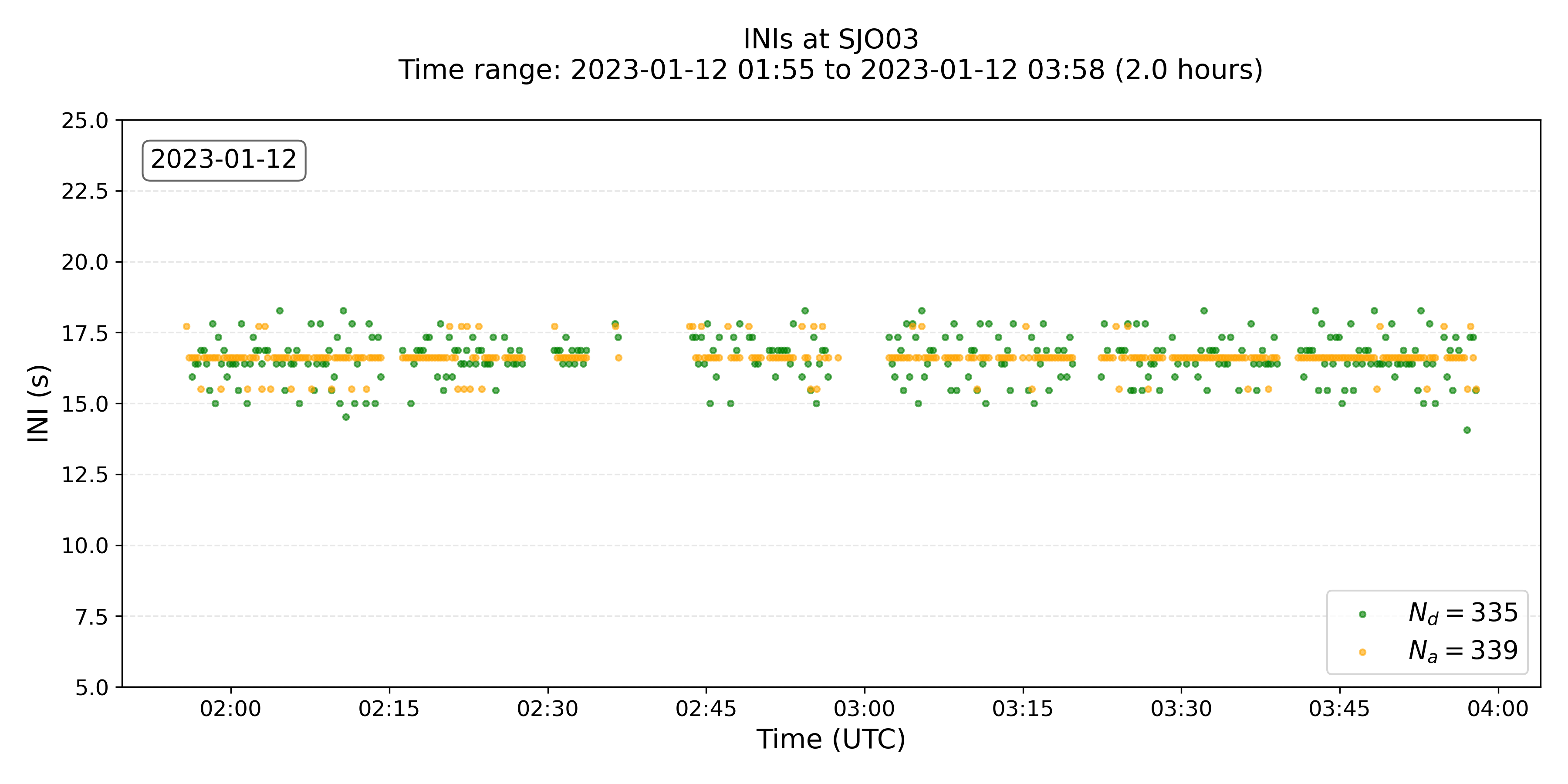}
        \put(-4,35){\large a)}              
    \end{overpic}
    
    \label{song1}
\end{subfigure}
\begin{subfigure}{\textwidth}
    \centering
    \begin{overpic}[trim=0 0 0 2cm,clip,width=.96\linewidth]{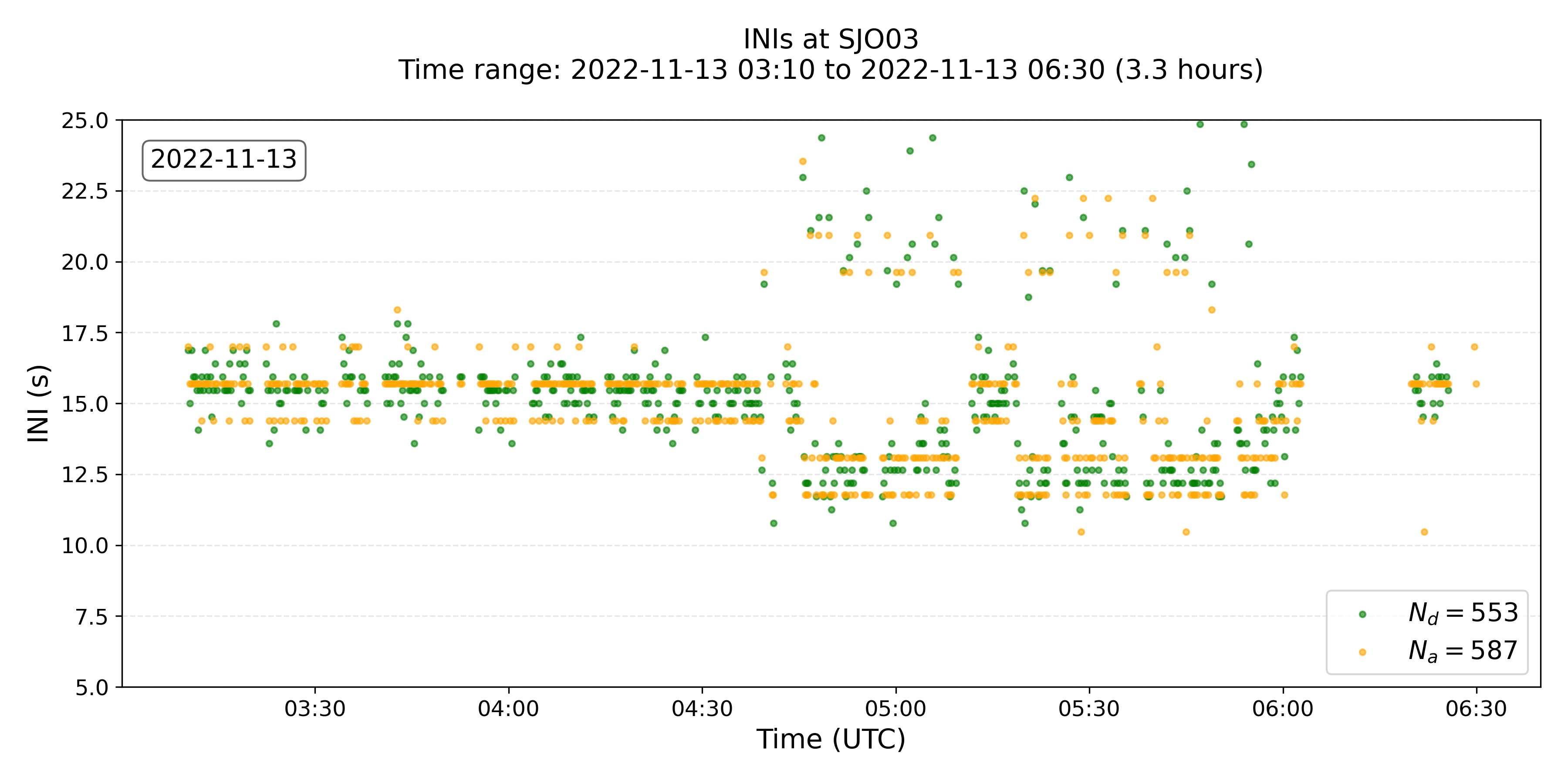}
        \put(-4,39){\large b)}             
    \end{overpic}

    \label{song2}
\end{subfigure}

\begin{subfigure}{\textwidth}
    \centering
    \begin{overpic}[trim=0 0 0 2cm,clip,width=.96\linewidth]{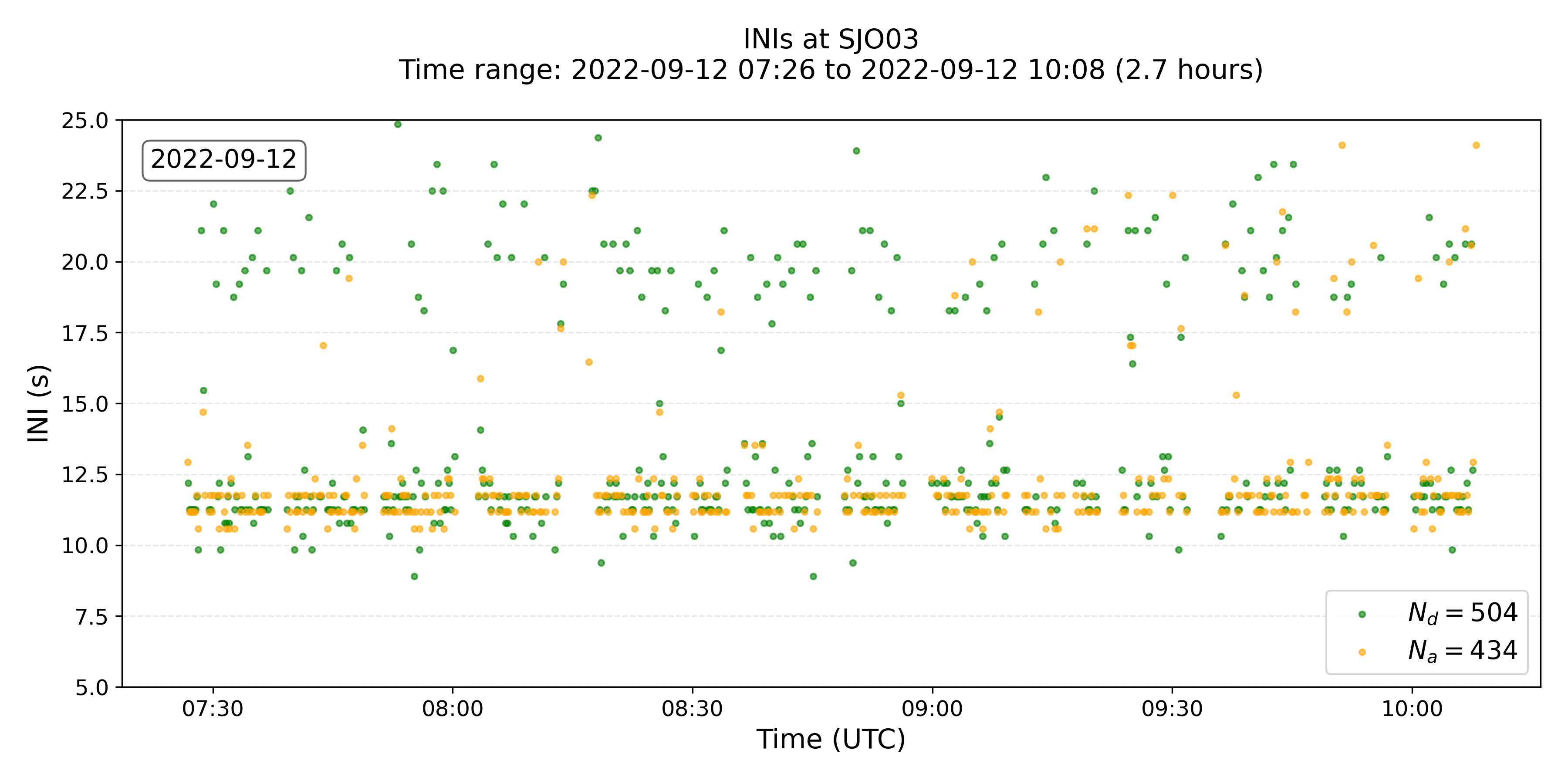}
        \put(-4,39){\large c)}             
    \end{overpic}
    \label{song3}
\end{subfigure}
\caption{Comparisons of INIs derived from model detections $N_{d}$ (green) and annotations $N_{a}$ (yellow). These songs show a \textbf{a)} singlet INI pattern centred at 16.5~s, \textbf{b)} doublet INI pattern alternating between 12.5~s and 16.5~s, and \textbf{c)} predominantly singlet INI pattern centred at 11.5~s.}
\label{fig:Song_INI_comparisons}
\end{figure}


\subsubsection*{Anomalous stations}

\begin{figure}[h!]
  \includegraphics[width=0.6\textwidth, trim={0 0 0 0}, clip]{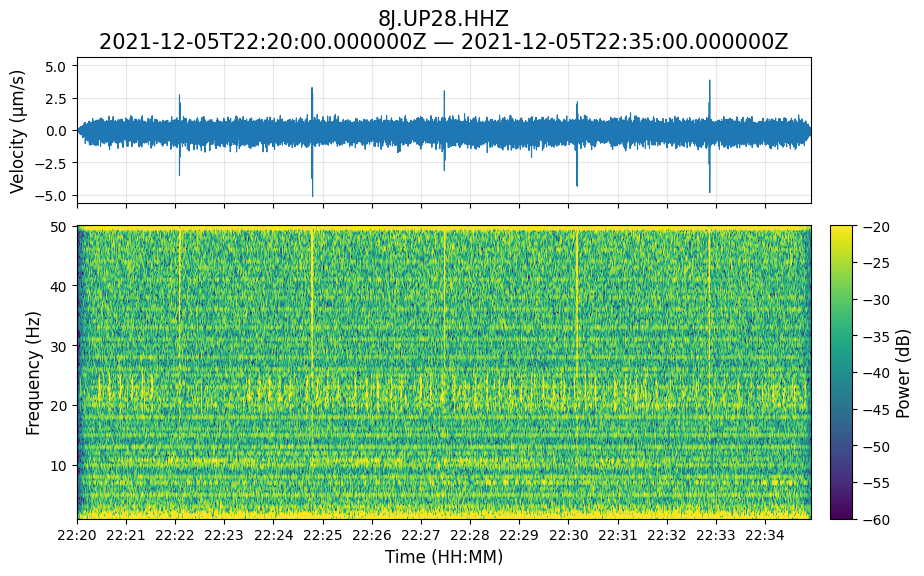}
  \caption{Example of anomalous noise signals in station UP28 masking fin whale vocalisations around $\sim$22~Hz. This noise appeared at multiple distinct frequency bands and was observed throughout the whole deployment.}
  \label{fig:anomaly_UP28}
\end{figure}

\begin{figure}[h!]
  \includegraphics[width=0.6\textwidth, trim={0 0 0 0}, clip]{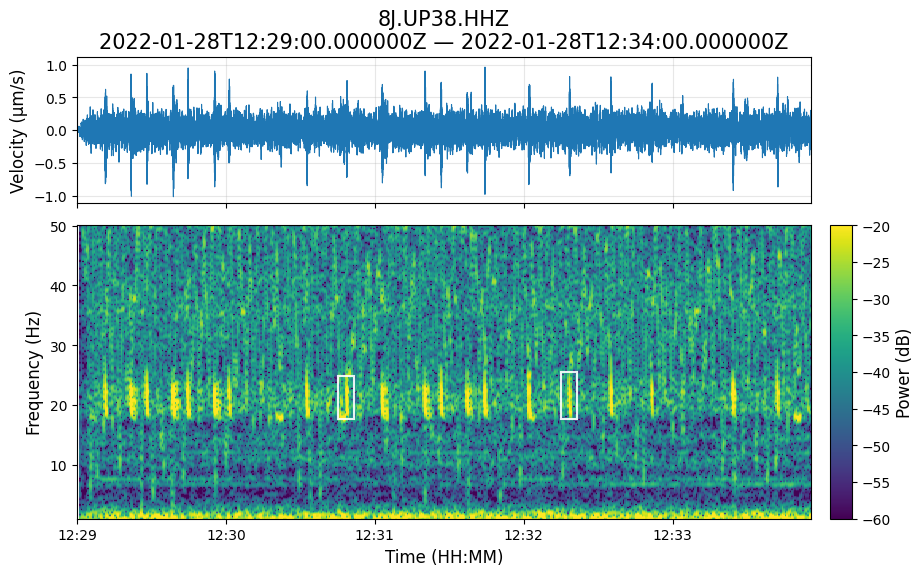}
  \caption{Broadband noise in station UP38 observed throughout the deployment, which may have affected the detectability of calls (white boxes)}
  \label{fig:anomaly_UP38}
\end{figure}

\begin{figure}[h!]
  \includegraphics[width=0.6\textwidth, trim={0 0 0 0}, clip]{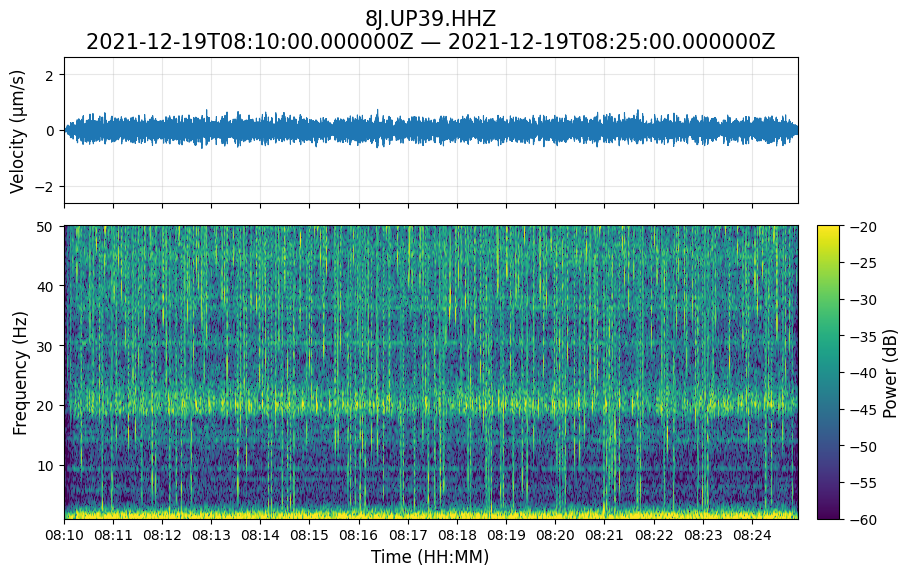}
  \caption{Broadband noise in station UP39, similar to those in station UP38.}
  \label{fig:anomaly_UP39}
\end{figure}


\subsubsection*{Other sources of masking}
\begin{figure}[htp!]
\centering

\begin{subfigure}{0.8\textwidth}
    \centering
    \begin{overpic}[trim=0 0.8cm 3cm 5.5cm,clip,width=\linewidth]{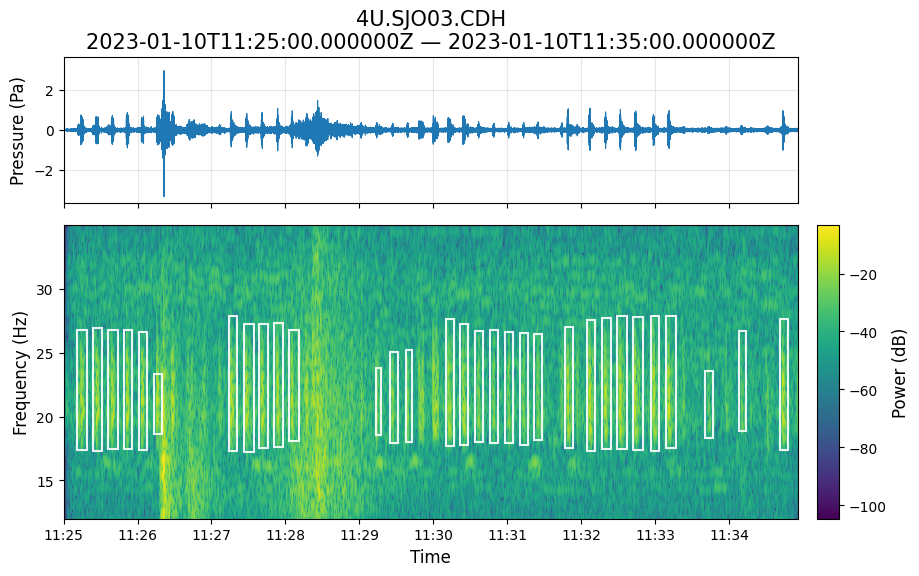}
        \put(-4,22){\large a)}              
    \end{overpic}

    \label{song1}
\end{subfigure}

\begin{subfigure}{0.81\textwidth}
    \centering
    \begin{overpic}[trim=0.3cm 0.8cm 3cm 5.5cm,clip,width=\linewidth]{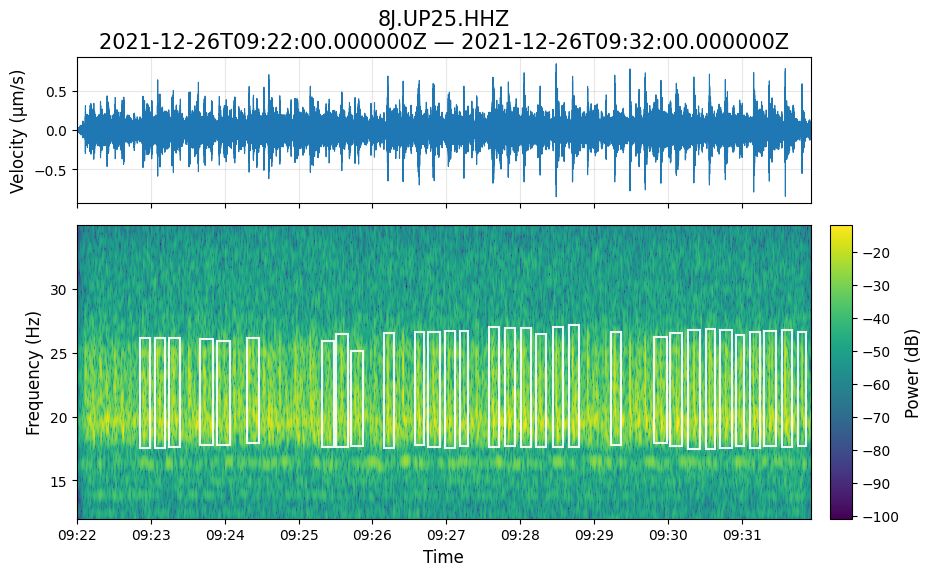}
        \put(-4,22){\large b)}             
    \end{overpic}

    \label{song2}
\end{subfigure}

\begin{subfigure}{0.8\textwidth}
    \centering
    \begin{overpic}[trim=0 0 3cm 5.5cm,clip,width=\linewidth]{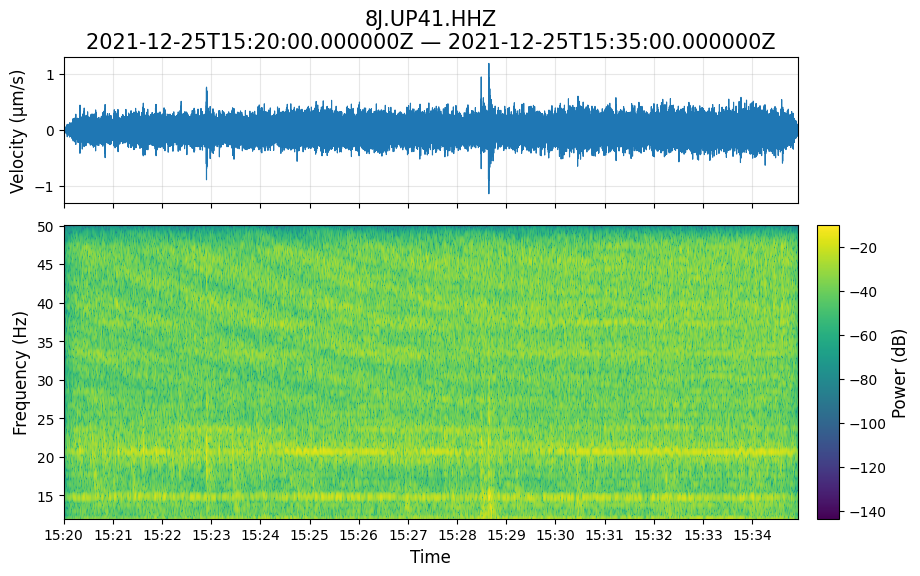}
        \put(-4,22){\large c)}             
    \end{overpic}
    \label{song3}
\end{subfigure}
\caption{Detected 20-Hz notes (white boxes) overlaid on spectrograms of OBS recordings illustrating sources of missed detections. \textbf{a)} Calls at station SJO03 masked by a regional earthquake, visible as a broadband transient signal. \textbf{b)} A dense chorusing event at station UP25 in which overlapping calls from multiple individuals reduce call separability. \textbf{c)} Masking due to shipping noise at station UP41.}
\label{fig: Spectrograms_masking_examples}
\end{figure}

\end{document}